\documentclass[11pt,a4]{article}
\usepackage[applemac]{inputenc}  
\usepackage{amssymb,amsmath,amsthm,amsfonts,amscd,amstext}
\usepackage[mathscr]{eucal}
\usepackage{array}
\usepackage{graphicx}
\newtheorem{lem}{Lemma}[section]
\newtheorem{prop}[lem]{Proposition}
\newtheorem{thm}[lem]{Theorem}

\newtheorem{df}{Definition}[section]

\newcommand{\Aut}{\mathrm{Aut}}
\def\aut{\operatorname {Aut}} 
\def\C{{\mathbb C}}  

\def\coker{\operatorname {coker}} 
\def\d  {{\rm d}}
\def\D{{\mathbb D}}

\def\der{\operatorname {Der}} 

\def\Det{\operatorname {Det}}
\def\Diff{\operatorname {Diff}}
\def\diff{\operatorname {diff}}

\def\Gr{\operatorname{Gr}}   
\def\H {\mathbb H} 

\def\id{\operatorname{id}}  

\def\ker{\operatorname{ker}}  

\def\OO{\cal O}  

\def\pr{\operatorname{pr}} 

\def\R{{\mathbb R}}    

\def\tr{\operatorname{tr}}  
\def\vect{\operatorname{Vect}}
\newcommand{\Vir}{\mathrm{Vir}}
\def\vir{\operatorname{Vir}} 
\def\Vir{\operatorname{Vir}} 

\def\Z{{\mathbb Z}}    
\def\N{{\mathbb N}}    

\def\res{\operatorname{Res}}  

\newcommand{\unit}{\mathbf{1}}

\newcommand{\Xope}{\mathfrak{O}}

\title{\bf The Global Geometry of Stochastic L{\oe}wner Evolutions}
\author
{Roland Friedrich\\
\\
\normalsize{MPI}\\
\normalsize{e-mail: rolandf@mpim-bonn.mpg.de}
}
\date{}
\begin{document}
\maketitle
\begin{abstract}
In this article we develop a concise description of the global geometry which is underlying the universal construction of all possible generalised Stochastic L{\oe}wner Evolutions. The main ingredient is the Universal Grassmannian of Sato-Segal-Wilson. We illustrate the situation in the case of univalent functions defined on the unit disc and the classical Schramm-L{\oe}wner stochastic differential equation. In particular we show how the Virasoro algebra acts on probability measures. This approach provides the natural connection with Conformal Field Theory and Integrable Systems. 
\end{abstract}
\tableofcontents
\section{Introduction}
The  aim of this contribution, which is primarily based on two talks, namely 2003 in Edinburgh and 2008 in Kyoto, is to convey  a rather comprehensive and concise picture of the framework in which one should perceive the L{\oe}wner equation, and in particular ``Stochastic L{\oe}wner Evolutions" (SLE). The immediate reward is to see how to connect SLE with Conformal Field Theory (CFT) but also with Integrable Systems, as arising from the KdV-equation. So, one naturally has to revert to the seminal paper~\cite{KNTY}. The fundamental idea of the works~\cite{F, Fth, FK, K}, which can be considered as the universal approach to (generalised) (S)LE, is based on the insight that SLE embeds naturally into the picture of Virasoro uniformisation~\cite{ADKP, KNTY, K1, BMS}. 

In the various parts of this text we give the necessary background from the general theory, as indicated in the Bibliography section, and apply it to derive plenty of new connections with the L{\oe}wner equation. In particular, we show that seemingly different classical approaches can be treated in a uniform manner at the analytical level by the theory of univalent functions.

A comment concerning the style of writing. We have tried to give a  readable account, first by seamlessly interweaving proper results, mainly in the introductory sections, into the  pre-existing mathematical fabric  and then utilising it in the later parts to work out the new theory.

\section{Univalent functions, the Grunsky matrix and Faber polynomials}
\subsection{Univalent functions and the Lie Group $\aut({\cal O})$}
An injective function $f$ in a domain of the complex plane is called {\bf univalent} (or {\bf schlicht}). Further, an analytic and injective function is called a {\bf conformal mapping}. The most important class of such functions is the set $S$ of functions $f$ which are analytic and injective in the unit disc $\D:=\{z\in\C~|~|z|<1\}$, normalised by the conditions $f(0)=0$ and $f'(0)=1$. Therefore every $f\in S$ has a power series expansion around the origin of the form:
$$
f(z)=z+a_2 z^2+a_3 z^3+\cdots,\qquad |z|<1.
$$
The most prominent example of a function of class $S$ is the {\bf Koebe function}
$$
k(z)=z(1-z)^{-2}=z+2z^2+3z^3+\cdots~,
$$
which maps the unit disc onto $\C\setminus (-\infty,-\frac{1}{4}]$. 

A class related to $S$ is the set of functions $\Sigma$, which are analytic and univalent in the domain $\D_{\infty}:=\{z\in\C~|~|z|>1\}$ exterior to $\D$ and which have a simple pole at infinity with residue 1. Therefore for an element $g\in\Sigma$, we have around $\infty$ the following series development 
$$
g(z)=z+b_0+b_1z^{-1}+b_2 z^{-2}+\cdots.
$$
Every such $g$ maps $\D_{\infty}$ onto the complement of a compact connected set. If one properly adjusts the constant term $b_0$ then the range will not contain the origin $0$. For each $f\in S$, the map
$$
f\mapsto g(z):=\frac{1}{f\left(\frac{1}{z}\right)}=z-a_2+(a_2^2-a_3)z^{-1}+\cdots~,
$$
called an {\bf inversion}, establishes a bijection with the set $\Sigma_0$ of functions $g$, not mapping onto $0$.

More generally, if we denote by ${\cal O}$ the complete topological $\C$–algebra $\C[[z]]$ of formal power series, and by $\Aut({\cal O})$ the group of continuous automorphisms of ${\cal O}$, then such a continuous  automorphism $\rho$ is determined by its action on the generator $z$ of $\C[[z]]$. It may be represented by a formal power series of the form $a_1z + a_2z^2 +\cdots,$ with $a_1\neq0$. The unique maximal ideal ${\frak m}$ of ${\cal O}$ is then given by the set $z {\cal O}=\left\{\sum_{n=1}^{\infty} a_n z^n\right\}$. Before we describe the structure of $\Aut({\cal O})$ in more detail, let us introduce the following spaces of formal power series. 
\begin{eqnarray*}
\Aut({\cal O}) & := & \left\{f(z)=\sum_{n=1}^{\infty} a_n z^n~; a_1\neq 0\right\}~, \\
\Aut_+({\cal O}) & := & \left\{f(z)=z+\sum_{n=2}^{\infty} a_n z^n~ \right\}~, \\ 
\Aut_0({\cal O}) & := & \left\{f(z)=z+\sum_{n=3}^{\infty} a_n z^n~ \right\}~,
\end{eqnarray*} 
and their counterparts at infinity, i.e.
\begin{eqnarray*}
\Aut({\cal O}_{\infty}) & := & \left\{g(z)=b z+\sum_{n=0}^{\infty} b_n z^{-n}~; b\neq 0\right\}~, \\
\Aut_+({\cal O}_{\infty}) & := & \left\{g(z)=z+\sum_{n=0}^{\infty} b_n z^{-n} \right\}~, \\ 
\Aut_0({\cal O}_\infty) & := & \left\{g(z)=z+\sum_{n=1}^{\infty} b_n z^{-n} \right\}~. 
\end{eqnarray*} 
These power series should be understood as the completions of the spaces of analytic functions which are locally univalent around either the origin or infinity. The completion is with respect to the natural filtration  induced by the unique maximal ideal. So, these spaces are naturally seen to be Lie groups and we have the following Lie groups and Lie algebras:
\begin{eqnarray*}
\Aut_+({\OO}) &  &\qquad\der_+({\OO})=z^2\C[[z]]\partial_z \\
\cap\qquad & & \qquad\qquad\cap \\
\Aut({\OO}) & &\qquad\der_0({\OO})=z\C[[z]]\partial_z \\
& & \qquad\qquad\cap \\
& & \qquad\der({\OO})=\C[[z]]\partial_z
\end{eqnarray*}
whose properties are summarised in
\begin{prop}[\cite{KNTY}]
\begin{enumerate}
  \item $\aut({\cal O})$ acts on itself by composition, and it is a semi-direct product of $\C^*$ and $\Aut_+({\OO})$.
 \item $\aut({\cal O})_+$ is a pro-algebraic group, i.e. $\aut({\cal O})_+=\varprojlim \aut_+({\cal O}/\mathfrak{m}^n)$.
  \item $\operatorname{Lie}(\Aut({\cal O}))=\der_0({\cal O})$, $\operatorname{Lie}(\Aut_+({\cal O}))=\der_+({\cal O})$, and the exponential map $\exp:\der_+({\OO})\rightarrow\Aut_+({\OO})$, is an isomorphism. \end{enumerate}
\end{prop}
We have a natural embedding
$$
\aut({\cal O})\hookrightarrow \C^{\N},\qquad \rho\mapsto (a_1, a_2,\dots)
$$
and equivalently for the other Lie groups from above. 

In particular, we see that $\Aut_+({\OO})$ has the structure of an affine space $f_0+z^2\C[[z]]$ where $f_0(z)\equiv z$. The set  $\{c_1, c_2,\dots\}$, where $c_n:=a_{n+1}$, yields affine co-ordinates.

Now, it follows from the {\bf Bieberbach-de~Branges~Theorem}, that its affine coefficients $c_n$, satisfy 
$$
|c_n|\leq n+1\qquad \forall n\geq1~,
$$ 
so the set of schlicht functions $S$, can be identified with a subset of 
$$
\prod_{n\geq 1} \operatorname{Ball}_{\C}(0,n+1)~, 
$$
and therefore is  a (contractible) manifold. Let us note that the above condition is necessary but not sufficient for a function $f$ to be in the class $S$. 

Another suitable but more restrictive topology would be the usual Fréchet topology on the vector space of holomorphic functions. 

Finally, let us denote by ${\cal M}_0\subset\Aut{(\OO)}$, the subset of analytic elements, i.e., whose power series converges at least in a neighbourhood of $0$; and correspondingly ${\cal M}_{\infty}\subset\Aut({\OO}_{\infty})$, and ${\cal M}$ stands for any of the two, which should follow from the context. 
\subsection{The Grunsky matrix and Faber polynomials}
A necessary and sufficient condition for a function defined on the unit disc to be univalent was given by Grunsky, via a system of inequalities, which are called the {\bf Grunsky inequalities}. 

We shall now briefly recall the definitions of Faber polynomials and Grunsky coefficients~\cite{Pom, DSch}.

Let us consider a {\bf complementary pair} $(f,g)$ of univalent functions which map onto non-overlapping  regions of the Riemann sphere. This means that $f(z)=  a_1z +a_2z^2+\cdots~,$ with $r=|a_1|$ (and after a rotation we may assume $a_1>0$),  is analytic and univalent in $|z|<1$, and 
$$
g(w)=Rw+b_0+b_1w^{-1}+b_2 w^{-2}+\cdots,\quad R>0,
$$
is analytic and univalent in $|w|>1$. Further, suppose that $f$ and $g$ have disjoint ranges, i.e., $f(z)\neq g(w)$ for $|z|<1$ and $|w|>1$. 

Then the following three matrices  $[c_{nm}]$, $[d_{nm}]$ and  $[e_{nm}]$ can be formed by the expansions
\begin{eqnarray}
\label{Grunsky}
\log\frac{f(z)-f(w)}{z-w} & = & -\sum_{m=0}^{\infty}\sum_{n=0}^{\infty}\, c_{mn}\, z^m w^n,\qquad |z|<1,~|w|<1;\\\nonumber
\log\frac{g(z)-g(w)}{z-w} & = & -\sum_{m=0}^{\infty}\sum_{n=0}^{\infty}\, d_{mn}\, z^{-m} w^{-n},\qquad |z|>1,~|w|>1;\\
\log\frac{f(z)-g(w)}{z-w} & =&-\sum_{m=0}^{\infty}\sum_{n=0}^{\infty}\, e_{mn}\, z^m w^{-n},\qquad |z|<1,~|w|>1.\nonumber
\end{eqnarray}

One observes that the matrices $[c_{mn}]$ and $[d_{mn}]$ are symmetric,  i.e., $c_{mn}=c_{nm}$ respectively  $d_{mn}=d_{nm}$,  and that $c_{00}= -\log r$, $d_{00} = -\log R$ and 
$e_{00} =  -\log R$, holds.
\begin{df}
The matrices $[c_{mn}]$, $[d_{mn}]$ and  $[e_{mn}]$, are called the {\bf Grunsky matrices} and the coefficients $c_{mn}$, $d_{mn}$ and $e_{mn}$, the {\bf Grunsky coefficients}.
\end{df}
Let us unify the notation of the coefficients under the assumption that $R=1$. Set 
$b_{-m,-n}:=c_{mn}$ for $m,n\geq0$ and $b_{mn}=d_{mn}$ for $m,n\geq1$.

Then for $n\geq1$ the {\bf Faber polynomials} $G_n(w)$ for general $g$, respectively $F_n(w)$ for $f$ with  are defined by
\begin{eqnarray}
\label{Faber}
\log\frac{g(z)-w}{Rz} & = & -\sum_{n=1}^{\infty}\,\frac{G_n(w)}{n} z^{-n}, \\\nonumber
\log\frac{w-f(z)}{w} & = &\log\frac{f(z)}{rz} -\sum_{n=1}^{\infty}\,\frac{F_n (w)}{n} z^{n},  
\end{eqnarray}
where $G_n(w)$ is a polynomial of degree $n$ in $w$ and $F_n(w)$ is a polynomial in $1/w$ of degree $n$.
The Faber polynomials and the Grunsky coefficients can be related by the following identities (for $R=r=1$): 
\begin{eqnarray}
\label{Tao2.4}
G_{n}(g(w))=w^n+n\sum^{\infty}_{m=1} b_{nm} w^{-m}, &  & G_{n}(f(w))=nb_{n,0}+n\sum^{\infty}_{m=1} b_{n,-m} w^{-m}, \\
F_{n}(g(w))=-nb_{-n,0}+n\sum^{\infty}_{m=1} b_{m,-n} w^{-m}, &  & F_{n}(f(w))=w^{-n}+n\sum^{\infty}_{m=1} b_{-n,-m} w^{m}. 
\end{eqnarray}
\subsection{$\Diff(S^1)/S^1$ and Kirillov's Kähler manifold of univalent functions}
The dense subclass $S_K\subset S$, consisting of univalent functions $f$ which extend to a smooth injective function on the boundary $\partial\D$, and satisfy the additional condition $f'(e^{it})\neq0$, was first considered by A.A.~Kirillov. The importance of $S_K$ comes from the following facts. 

Let us denote by $\Diff_+(S^1)$, the infinite dimensional Lie group of $C^{\infty}$, orientation preserving diffeomorphisms of the circle $S^1$ and, by abuse of notation, the subgroup of rotations $\operatorname{Rot}_+(S^1)$ also by $S^1$. This space has been extensively studied in the litterature, both in mathematics and physics~\cite{BR, KY, M, NS, N, PS} (and references therein). Then, as can be shown by conformal welding and a theorem of J.~Moser, there exists a canonical bijection~\cite{KY}
\begin{equation}
\label{Kirillov}
K:S_K\rightarrow\Diff_+(S^1)/S^1~.
\end{equation}
This implies that $S_K$ is a homogenous space under the left action of $\Diff_+(S^1)$, and that the identity function $f_0(z)\equiv z$, is the stabiliser of the left action of $S^1$. 

The corresponding Lie algebra $\diff(S^1)$ can be identified with the $C^{\infty}$, real-valued  left-invariant vector fields $v(t)d/dt$, with the Lie bracket given by
$$ 
[\varphi, \psi]=\varphi\dot{\psi} -\dot{\varphi}\psi,\quad \varphi,\psi\in\diff(S^1)~.
$$ 
The Lie algebra $\diff(S^1)$, has a natural basis, with respect to the Fréchet topology, consisting of trigonometric polynomials
\begin{equation}
\label{TrigPoly}
\varphi_m(t) := \cos(m t),\quad \psi_n(t) :=\sin(nt),\quad m=0,1,2,\dots, ~ n= 1, 2,\dots ~.  
\end{equation}
Let us fix two parameters $c,h\in\R$. Then one can define a bilinear antisymmetric form on $\diff(S^1)$:
\begin{equation}
\label{ExtGFcocycle}
\omega_{c,h}(v_1,v_2):=\frac{1}{2\pi}\int_0^{2\pi}\left((2h-\frac{c}{12})v_1'(t)-\frac{c}{12}v_1'''(t)\right)v_2(t)\,dt~.
\end{equation}

Let us remark that the classical {\bf Gelfand-Fuks cocycle} corresponds to $\omega_{1,0}$. To have convergence in~(\ref{ExtGFcocycle}), the vector fields should be of Sobolev class $3/2$, (cf.~\cite{AMT, NS, SchTZ}).

Let us now consider the complexification of the Lie algebra $\diff(S^1)$, namely
$$
\diff_{\C}(S^1):=\diff(S^1)\otimes_{\R}\C~.
$$
It has a topological basis $\{e_k\}_{k\in\Z}$, given by
\begin{equation}
\label{Vir_Top_base}
e_k:=-ie^{ikt}\frac{d}{dt}~,
\end{equation}
which satisfies the commutation relations of the Witt algebra, i.e.,
$$
[e_m, e_n]=(n-m)\, e_{m+n}~.
$$
Now, the complex Virasoro algebra $\vir_{\C}$ is spanned by the polynomial vector fields $e_k$, and an additional element $\mathfrak{c}$. They satisfy the following commutation relations: 
\begin{eqnarray*}
[{\mathfrak{c}},e_n] & = & 0~,\\
{[e_m, e_n ]} & = & { [e_m, e_n ]}+\omega_{c,h}(e_m,e_n)\cdot{\mathfrak{c}}~, 
\end{eqnarray*}
where $\omega_{c,h}$, is the cocycle given by~(\ref{ExtGFcocycle}). In Physics the numbers $c$ and $h$ are usually called the {\bf central charge} resp. the {\bf highest weight}.

Then, on the set of vector fields with zero mean, i.e. 
$$
\diff'(S^1):=\{ v\in\diff(S^1)~|~\int_0^{2\pi} v(t) dt=0 \}
$$
one can introduce a complex structure. Namely, the Schwarz kernel defines a familiar integral operator $I$, called the {\bf Hilbert transform}, on $L^2(S^1)$ by
\begin{equation}
\label{Hilbert_trafo}
(If)(t):=\frac{1}{\pi}\, pv. \int_0^{2\pi}\cot\left(\frac{t-\phi}{2}\right) f(\phi)\, d\phi~.
\end{equation}
As the integral diverges, it has to be regularised by taking its principal value
$$
If(t):=\frac{1}{\pi}\lim_{\epsilon\to0} \left(\int_0^{t-\epsilon}+\int_{t+\epsilon}^{2\pi}\right)~.
$$
A calculation shows that
$$
I(e^{int})=\begin{cases}
      i e^{int}& n>0, \\
      0 & n=0,\\
      -i e^{int} & n<0,
\end{cases}
$$
which yields $I^2=-\id$. Further, the operator $I$ can be extended by continuity to a bounded operator on the entire space. Therefore, the Hilbert transform defines a canonical almost-complex structure $J$, on $\diff'(S^1)$, which can be equivalently written as:
\begin{equation}
\label{M1.2.5}
J(v(t)):=\sum_{k=1}^{\infty}\left(-a_k\sin(k t)+b_k\cos(kt)\right)~.
\end{equation}
Let us note, that the extended complex structure on the complexified tangent space allows us to go from tangent vector fields to normal vector fields on the circle.

If we denote by $T\Aut_+({\cal O})$ the real tangent space, i.e. the vector space of real first order differential operators on $\Aut_+({\cal O})$, then the above complex structure $J$ on $\Aut_+({\cal O})$ induces the splitting
$$
T\Aut_+({\cal O})\otimes_{\R}\C=T^{(1,0)}\Aut_+({\cal O})\oplus T^{(0,1)}\Aut_+({\cal O})~.
$$

Now, by the Kirillov identification one can define a two-parameter family of Kähler metrics on $S_0$, by putting $w_{c,h}:=\omega_{c,h}(\cdot, J(\cdot))$. At the origin $f_0\equiv z$, in the affine co-ordinates $\{c_k\}$, it has the form~\cite{KY}
\begin{equation}
\label{Kaehler_m}
w_{c,h}:=\sum_{k=1}^{\infty}\left(2hk+\frac{c}{12}(k^3-k)\right)\,dc_k\wedge d\bar{c}_k~.
\end{equation}
This metrics also  generate the Weil-Petersson metric. 

Let us conclude this section by summarising the various analytic subspaces of the space of formal power series and their additional structures:  
\[
\begin{CD}
\label{Kir_Aut(O)}
\Diff_+(S^1)/S^1@>\text{Kirillov}>\text{bijection}>S_0@>\text{dense}>>S@>>>\Aut_+({\cal O})@>>>\Aut({\cal O})~.
\end{CD}
\]

\section{Subordination chains and the Neretin-Segal semi-group}
\begin{figure}[htbp!]
\begin{center}
\includegraphics[scale=0.4]{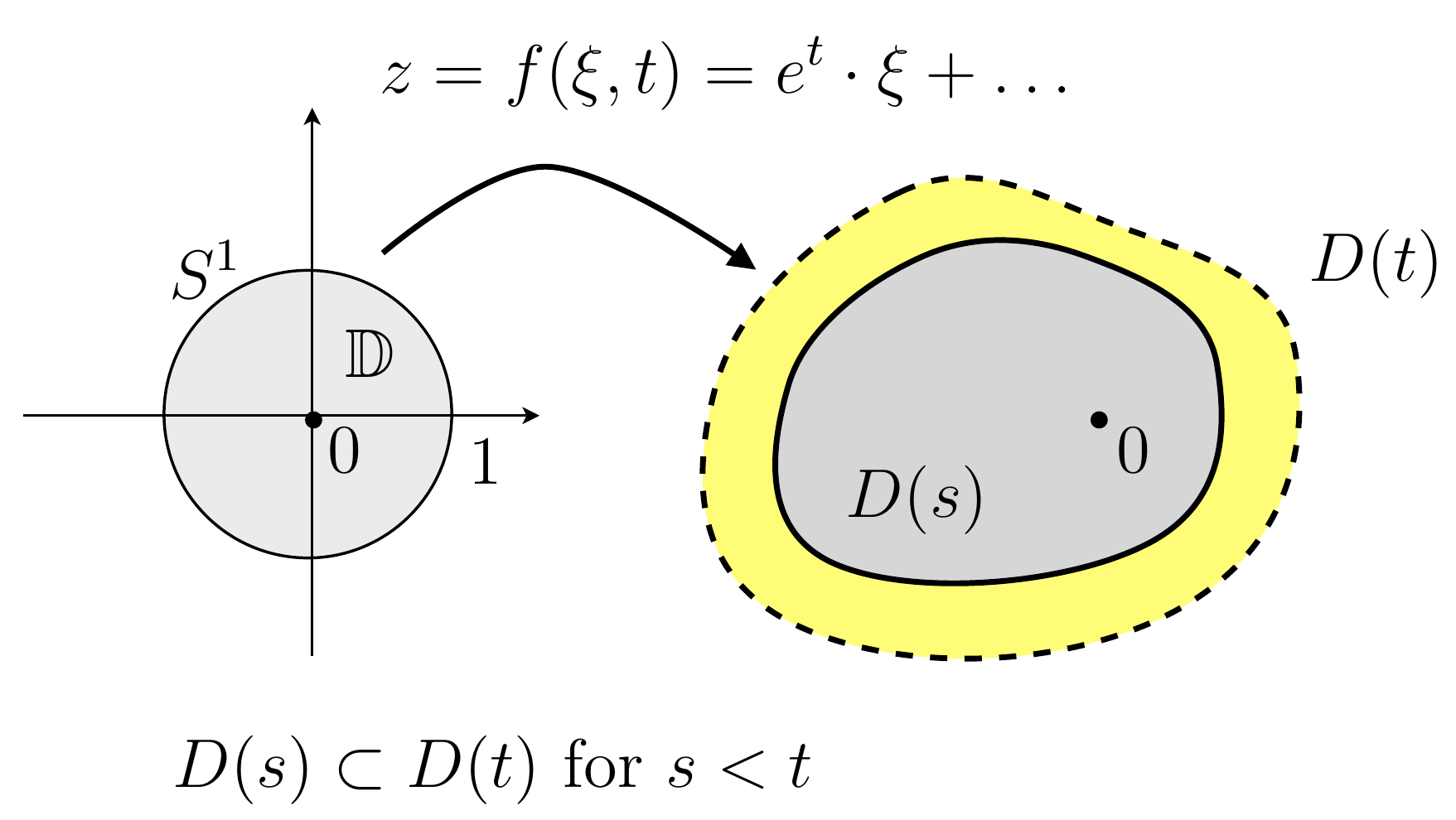}
\caption{Schematic representation of a subordination chain.} 
\label{SubChain}
\end{center}
\end{figure}

Let us start with the following important (cf. Fig~\ref{SubChain}),
\begin{df}
A family of conformal maps $f_t:\D\rightarrow\C$ is called a {\bf subordination chain} if the following conditions hold:
\begin{itemize}
\item $\forall s,t\in\R_+~\text{and}~s\leq t: f_s(\D)\subset f_t(\D)$
\item $\forall t: f_t(0)=0$ and $f'_t(0)=e^t$
\item $f_0\equiv f$
\end{itemize}
\end{df}
We have to remark, that one could use instead of $0$ another base point for the family of domains and second, instead of increasing domains  one could have a family of strictly decreasing domains with the derivative at the origin being $f'_t(0)=e^{-t}$. In particular this is the standard parametrisation for the radial L{\oe}wner equation in the unit disc. 

Given a subordination chain of domains defined for $t\in[0,T)$, with $T\in\R_+\cup\{\infty\}$, by a theorem of Ch.~Pommerenke~\cite{Pom}, there exists a regular analytic function on $\D$,
\begin{equation}
\label{LKP_funct}
p(z,t)=1+p_1(t)z+p_2(t)z^2+\cdots,
\end{equation}
such that $\Re(p(z,t))>0$, i.e. maps the unit disc onto the right half-plane, and satisfies the partial differential equation, called the {\bf L{\oe}wner-Kufarev equation}
\begin{equation}
\label{Pommerenke}
\frac{\partial f}{\partial t}(z,t)=z\cdot\frac{\partial f}{\partial z}(z,t)\cdot p(z,t)~,
\end{equation}
for $z\in\D$ and for almost all $t\in[0,T)$.

The function $w_t:\D\rightarrow\D$, $w_t:=f_t^{-1}\circ f_0$ is called the {\bf transition function} of the subordination chain, and satisfies
$$
\lim_{t\to\infty} e^t w_t=f_0,\quad~\text{and}\quad~w_0\equiv\id_{\D}~.
$$
If we assume that the maps $f_t$ are smooth and injective up to the boundary, $w_t(S^1)$ gives a family of Jordan curves surrounding the origin. 
Hence, the transition functions can be seen as elements of the {\bf Neretin-Segal semigroup}~\cite{N, PS} which has different but equivalent realisations. In our context the following will do.
The {\bf Neretin semi-group} is the set of analytic maps $\rho:S^1\rightarrow\D$ such that
\begin{enumerate}
  \item $\rho'(e^{i\theta})\neq0$,
  \item $\rho(S^1)$ is a smooth Jordan curve, parametrised counter-clockwise.
\end{enumerate}

The tangent cone $\mathfrak{ner}$ to the Neretin semigroup at the identity is the collection of inward-pointing vector fields on the unit circle. $\mathfrak{ner}$ is an open convex $\Diff_+(S^1)$-invariant cone in the Lie algebra $\vect_{\C}(S^1)$. Therefore, the Neretin semi-group can be  considered as a ``substitute" for the non-existing complex Virasoro group (cf. \cite{N, PS}).

\section{Boundary Variations and the Witt algebra}
In this section we shall first recall a derivation of the L{\oe}wner equation as done in~\cite{Fth}, based on the variational method of M.A. Lawrentjew und B.W. Schabat~\cite{LSch}. 

Let us be given a Jordan curve $C$ that bounds a domain $D=D(C)$. Further we assume that the map $f_C$ which maps $D(C)$ onto the unit disc $\D$, such that for an interior point $z_0\in D(C)$, it satisfies $f_C(z_0)=0$ and $f'_C(z_0)>0$, is known.

So, if $\widetilde{C}$ is a ``nearby" Jordan curve that also encircles the point $z_0$, we seek to know the function $f_{\widetilde{C}}$ that maps the domain $D(\widetilde{C})$ bounded by $\widetilde{C}$, again onto $\D$ and satisfies $f_{\widetilde{C}}(z_0)=0$, $f'_{\widetilde{C}}(z_0)>0$.
\begin{figure}[htbp!]
\begin{center}
\includegraphics[scale=0.4]{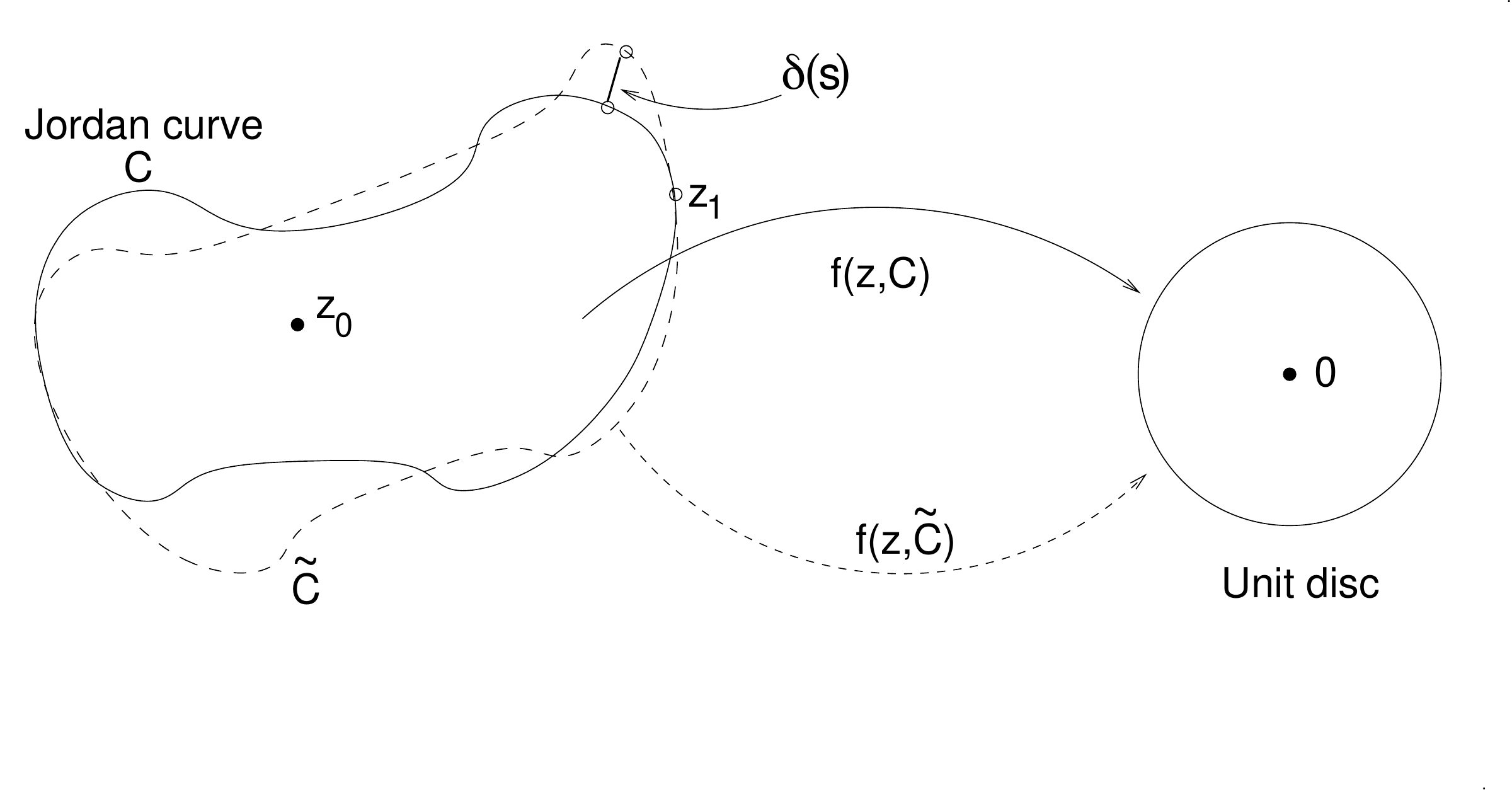}
\caption{Schematic representation of a boundary variation of a Jordan domain with the corresponding variation of the mapping function.} 
\label{BVariation}
\end{center}
\end{figure}
This problem can be solved in the following way. Let $z_1\in C$ be the point that is mapped by $f_C$ onto $1\in S^1\equiv\partial\D$, and let $s$ denote the arc length parmeter of $C$, measured from $z_1$ in the positive sense. Then $s$ ranges over the interval $[0, l]$, where $l$ denotes the length of $C$, as the point $\zeta$ traces over $C$ in the positive sense. Since we know $f_C$, we also know the dependence of the points $w=e^{i\theta}$ on the arc length $s$:
\begin{equation}
\label{arclength}
\frac{d\theta}{ds}=|f'_C(\zeta(s))|,\qquad\theta(s)=\int^s_0 |f'_C(\zeta(s))| ds.
\end{equation}
The curve $\widetilde{C}$ shall differ only slightly from  $C$ which means  that if $\delta(s)$ denotes the length of the segment of the normal of $C$
at the point $s$ between $C$ and $\widetilde{C}$ (cf. Fig.~\ref{BVariation}) then (we assume that $C$ is sufficiently smooth)
\begin{displaymath}
|\delta(s)|<\varepsilon,\quad |\delta'(s)|<\varepsilon, \quad |\delta''(s)|<\varepsilon, 
\end{displaymath} 
where $\varepsilon$ is a fixed and sufficiently small real number.
Further we shall adopt the convention that $\delta(s)>0$ if the segment is contained in $D(C)$, and $\delta(s)<0$ if it is outside of $D(C)$.  Under $f_C$ the curve $\widetilde{C}$ is mapped onto a curve $\widetilde{C}^*$ that differs only by little from $S^1$ and this new curve satisfies in polar coordinates $\rho e^{i\theta}$, up to an error term of higher order, the following approximate relation 
\begin{displaymath}
\rho\approx 1-|f'_C(\zeta(s))|\cdot\delta(s)=1-\frac{d\theta}{ds}\,\delta(s)=:1-\delta^*(s)
\end{displaymath}
where $s=s(\theta)$ is now a function of $\theta$, as it can be obtained from Eq.~(\ref{arclength}). 

Then, as explained in \cite{LSch}, the uniformising map of the domain $D(\widetilde{C})$ onto the unit disc is given by
\begin{equation}
\label{Schabat_445_6}
f_{\widetilde{C}}(z) = f_C(z)\left\{1+\frac{1}{2\pi}\int_0^{2\pi}\frac{e^{i\theta}+f_C(z)}{e^{i\theta}-f_C(z)}\,\delta^{\ast}(\theta)\, d\theta\right\} + O(\varepsilon^2)~.
\end{equation}

Now, if choose an arbitrary continuous family of Jordan curves which interpolates between the initial curve $C=C_0$ and a final curve $C_{t_f}$, such that the sequence of corresponding domains $D(C_t)$ forms a decreasing subordination chain, then we can derive a differential equation for the family of uniformising  maps $f_{C_t}$. 

Let us  denote the difference in the normal direction of two nearby curves $C_{t_1}$ and $C_{t_2}$ with $t_1\leq t_2$  as $\delta(t_1, t_2,s_{t_1})$,  where $s_{t_1}$ is the arc length parameter on $C_{t_1}$. Then the  corresponding quantity $\delta^*(t_1, t_2, \theta)$ is always positive, since the normal directions are always inward pointing, which follows from our assumption on how the contours grow. 

It follows from Eq.~(\ref{Schabat_445_6}), that to calculate the difference quotient 
\begin{displaymath}
\frac{f_{C_{t_2}}-f_{C_{t_1}}}{t_2-t_1}
\end{displaymath}
and then the differential, we have to consider the following expression 
\begin{equation}
\label{density}
\lim_{t_2\rightarrow t_1}\frac{\delta^*(t_1, t_2,\theta)}{t_2-t_1}=:\nu_{t_1}(\theta)
\end{equation}
where we assume, that this limit exists. Basically, (\ref{density}) is telling us, that we can differentiate under the integral in (\ref{Schabat_445_6}), and $\nu_t(\theta)$ is the derivative of $\rho(t_1, \theta)$, i.e. of the radial change. The quantity 
\begin{displaymath}
\nu_{t_1}(\theta) d\theta
\end{displaymath}
is therefore a Borel measure on the unit circle and the differential equation 
for a continuous deformation with ``densities" $\nu_t(\theta)$ is hence
\begin{equation}
\label{radial_loewner}
\frac{\partial f_{C_t}(z)}{\partial t}= f_{C_t}(z)\cdot\frac{1}{2\pi}\int_0^{2\pi}\frac{e^{i\theta}+f_{C_t}(z)}{e^{i\theta}-f_{C_t}(z)}\;\nu_t(\theta)\, d\theta~.
\end{equation} 
Let us also give the version for the inverse function $f_t:=f^{-1}_{C_t}$, which describes the evolution of the family of maps from $\D$ onto $D(C_t)$. Then equation~(\ref{radial_loewner}) becomes, 
\begin{equation}
\label{loew1}
 \frac{\partial f_t}{\partial t}(z)=- z f'(z)\cdot\frac{1}{2\pi}\int_0^{2\pi}\frac{e^{i\theta}+z}{e^{i\theta}-z}\cdot\nu_t(\theta)\, d\theta~.
\end{equation}
Let us first note that the integral in equation~(\ref{loew1}) is exactly the {\bf Herglotz representation} of the L{\oe}wner-Kufarev transition function $p(t,z)$ (cf. eq.~(\ref{LKP_funct})). Second, if one takes  for $\nu_t(\theta)$ a time-depended Dirac measure on the unit circle, i.e., $\delta_{x(t)}$ with $|x(t)|=1$, then for $e^{iu(t)}\equiv x(t)$, one obtains 
\begin{equation}
\label{Loewner_radial2}
\frac{\partial f_t}{\partial t}(z)=- z f'(z)\cdot\frac{e^{iu(t)}+z}{e^{iu(t)}-z}~,
\end{equation}
which is usually called the {\bf radial L{\oe}wner equation}. 
\subsection{Lie vector fields and Witt generators}
Here we shall derive the infinitesimal action of a boundary variation on the space $\Aut({\cal O})$, (cf.~\cite{Fth}). Further we shall give an explicit representation of the positive part of the Witt algebra in terms of the Lie vector fields, written in the affine co-ordinates on $S_K$. The expressions we shall obtain are identical to those obtained by A.A.~Kirillov and D.~Yur'ev~\cite{KY}, using another variational formula, namely the one by G.M.~Goluzin and M.~Schiffer. Besides boundary variations there are also the interior variations, which come from meromorphic functions. However, one can relate the two for smooth enough boundaries. 

Let  $v$ be a real $2\pi$-periodic function which can be written as a Fourier series:
$$
v(t)=a_0+a_1\cos(t)+b_1 \sin(t)+a_2\cos(2t)+b_2\sin(2t)+\cdots~.
$$
We shall use the following relations for $n\geq1$:
$$
\frac{1}{2\pi}\int_0^{2 \pi}\cos(nt)\,\frac{e^{it}+z}{e^{it}-z}\, dt=z^n~\quad\text{and}~\quad \frac{1}{2\pi}\int_0^{2 \pi}\sin(nt)\,\frac{e^{it}+z}{e^{it}-z}\, dt=-i z^n~.
$$

Therefore, we have according to expression~(\ref{loew1}) the following natural action of a boundary variation:
to the normal vector field $v(e^{it}){\bf n}$ on $S^1$ corresponds the Lie field, (which at the identity is an element of $\der_0({\cal O})$):
\begin{equation}
\label{LieAction}
({\cal L}_v(f))(z)=-z f'(z)\frac{1}{2\pi}\int_0^{2\pi}\frac{e^{it}+z}{e^{it}-z}\, v(t)\, dt~.
\end{equation}
Let $v_k(t):=-i e^{ikt}$. Then we have
\begin{prop}
The nonnegative part of the Witt basis acts in terms of Lie vector fields  as 
$$
L_k:={\cal L}_k f(z)=z^{1+k}f'(z).
$$  If $f\in S\subset{\Aut}_+({\cal O})$, then in the infinite affine co-ordinate system $\{c_n\}$, the following representation holds
\begin{equation}
\label{ }
L_k=\partial_k+\sum_{n=1}^{\infty}(n+1) c_n\partial_{n+k}~,\qquad k\geq 1~.
\end{equation}
\end{prop}
Let us remark that the constant vector field $v_0\equiv -i\cdot 1$, which induces dilatations, leads out of the class $S$. Also, a similar co-ordinate representation exists for the group $\Aut({\cal O})$.

From equation~(\ref{LieAction}), and the L{\oe}wner-Kufarev equation~(\ref{Pommerenke}), we see that the time derivatives of the uniformisation maps of a subordination chain are given by the Lie fields induced by the normal vector fields governing the infinitesimal deformation of the boundaries of the domains. 

Let us give a different representation of eq.~(\ref{Loewner_radial2}). Namely, by expanding the Schwarz kernel into its Taylor series, 
$$
\frac{e^{it}+z}{e^{it}-z}=1+2\sum_{n=1}^{\infty} e^{-int}z^n~,
$$
we obtain:  
\begin{prop}
The radial L{\oe}wner equation can be written in terms of the Lie vector fields as
\begin{equation}
\label{radialLoewner_Lie}
\frac{\partial f_t}{\partial t}(z)={\cal L}_0f(z)+2\sum_{n=1}^{\infty}e^{-inu(t)}{\cal L}_n f(z)~,
\end{equation}
where $\exp(iu(t))$, is the pre-image of the tip of the slit, at time $t$.
\end{prop}
So what is the role of meromorphic vector fields? They correspond to interior variations and can be translated into boundary variations, where the normal shift is given by projecting the difference vector onto the normal direction, i.e., 
$$
\delta{\bf n}(s)=\Re\left({\frac{1}{i t'(s)}\,\frac{a\rho^2}{t(s)-z_0}}\right)~.
$$

We see, that from the point of view of boundary variations only the positive Fourier modes play a role.
\section{Schramm-L{\oe}wner Evolutions}
The inverse problem related to a sub-ordination chain is called the {\bf L{\oe}wner-Kufarev} problem. Namely, given an initial domain and a driving function $p(z,t)$, when does the solution represent a subordination chain of simply connected domains. O.~Schramm~\cite{Sch} posed the problem with Brownian motion as the driving function, and with the claim that the complement of the so-evolving domains with respect to the unit disc should describe in principle the conformaly invariant scaling limit of various two-dimensional lattice models. The analytic part of this inverse problem was solved by him together with S.~Rohde~\cite{RSch}.  

The idea of O.~Schramm was to describe the (probability) measure on the path space by looking at an observable, which is defined as the position of the tip of the appropriately parametrised random curve, after uniformisation of the complement of the arc with respect to the disc.  
His claim was, that the statistics describing the position of such points should be Gaussian, centred at the starting point on the boundary of the co-domain, with the variance as the only free parameter. This was based on the following assumptions: 
\begin{itemize}
 \item the random driving function $U_t$ should be a continuous process,
 \item the distribution of $U_t-U_s$, is a function only of $t-s$, i.e., the  	  increments are identically distributed,
  \item $U_t-U_s$, is independent of $U_r$, $0\leq r\leq s$, i.e. it has a Markovian property. 
\end{itemize}
This would imply, that $U_t=\mu t+\sqrt{\kappa} B_t$, for a standard Brownian motion $B_t$. In case that the distribution of the image of the tip after uniformisation is symmetric about the starting point, then this yields $\mu=0$, i.e. the process is driftless.

The  chordal SLE$_\kappa$ curve $\gamma$ in the upper half-plane $\H$ describes the growth of simple random curves emerging from the origin and aiming at infinity, as follows:
\begin{df}[Schramm-L{\oe}wner Equation]
\label{loewner_eq}
For $z\in\H$, $t\geq0$ define $g_t(z)$ by $g_0(z)=z$ and 
\begin{equation}
\frac{\partial g_t(z)}{\partial t}  = \frac{2}{ g_t(z) - W_t}
\label{lowner}.
\end{equation}
\end{df}
The maps $g_t$ are normalised 
such that $g_t (z) = z + o(1) $ when $z \to \infty$
and  $W_t:= \sqrt{\kappa}\,B_t$ where  $B_t(\omega)$ is the standard
one-dimensional Brownian motion, starting at 0 and with variance
$\kappa>0$. Given the initial point $g_0(z)=z$, the ordinary differential
equation (\ref{lowner}) is well defined until a random time
$\tau_z$ when the right-hand side in (\ref{lowner}) has a pole.
There are two sets of points that are of interest, namely the preimage of infinity $\tau^{-1}(\infty)$ and its complement. For those in the complement we define:
\begin{equation}
\label{Khull}
K_{t}:=\overline{\{z\in\H: \tau(z)<t\}}
\end{equation}
The family $(K_t)_{t\geq0}$, called  hulls, is an increasing family of compact sets in $\overline{\H}$ where $g_t$
is the uniformising map from $\H\setminus K_t$ onto $\H$. Further there exists a continuous process $(\gamma_t)_{t\geq0}$ with values in $\overline{\H}$ such that
$\H\setminus K_t$ is the unbounded connected component of
$\H\setminus\gamma[0,t]$ with probability one. This process is
the trace of the SLE${}_{\kappa}$ and it can be recovered from
$g_t$, and therefore from $W_t$, by
\begin{equation}
\gamma_t  =  \lim_{z\rightarrow W_t, z\in\H} g_t^{-1}(z)~.
\end{equation}
The constant $\kappa$ characterises the nature of the resulting curves.  For $0<\kappa\leq 4$, SLE${}_{\kappa}$ traces over simple curves, for $4<\kappa<8$ self-touching curves (curves with double points, but without crossing its past) and, finally, if $8\leq\kappa$ the trace becomes space filling. 

Now, for another simply connected domain $D$ with two boundary points $A,B\in\partial D$ the chordal $SLE_{\kappa}$ in $D$ from $A$ to $B$ is defined as
\begin{displaymath}
K_t(D_{A,B}):=h^{-1}(K_t(\H,0,\infty))
\end{displaymath}
where $K_t(\H,0,\infty)$ is the hull as in (\ref{Khull}) and $h$ is the conformal map from $D$ onto $\H$ with $h(A)=0$ and $h(B)=\infty$. It is very important to note that this definition is justified because of the assumed conformal invariance of the measures.

Let us rewrite~(\ref{lowner}) in It{\^o} form, by setting $f_t(z):=g_t(z)-W_t$, which now satisfies the stochastic differential equation
\begin{equation}
\label{Loewner-Ito}
df_t(z)=\frac{2}{f_t(z)} dt-dW_t~.
\end{equation}
For a non-singular boundary point $x\in\R$, we can read off the generator $\hat{A}$ for the It{\^o}-diffusion $X_t:=f_t(x)$ as 
$$
\hat{A}=2\frac{1}{x}\frac{d}{dx}+\frac{\kappa}{2}\frac{d^2}{dx^2}~.
$$
Defining the first order differential operators 
\begin{equation}
\label{Witt}
\ell_n:=-x^{n+1}\frac{d}{dx}\qquad n\in\Z~,
\end{equation}
we obtain 
$$
\hat{A}=\frac{\kappa}{2}\ell^2_{-1}-2\ell_{-2}~.
$$
Let us note, that the differential operators~(\ref{Witt}) form a representation of the Witt algebra. 
\section{The Universal Grassmannian and Determinant Line Bundles}
We are going to recall briefly the definition of the Segal-Wilson Grassmannian of a Hilbert space~\cite{PS}. This is a functional analytic realisation of M.~Sato's ``Universal Grassmannian Manifold" (UGM) (cf. e.g.~\cite{KNTY}). 

Let $H$ be an infinite-dimensional separable complex Hilbert space and let
$$
H=H_+\oplus H_-
$$ 
be an orthogonal decomposition into closed infinite dimensional subspaces $H_+$ and $H_-$. Then $\Gr_{SW}(H)\equiv\Gr(H)$, is the set of closed subspaces $W\subset H$ satisfying 
\begin{enumerate}
  \item the orthogonal projection $\pr_+:W\rightarrow H_+$ is a Fredholm operator,
  \item the orthogonal projection $\pr_-:W\rightarrow H_-$ is a Hilbert-Schmidt operator.
\end{enumerate}
$\Gr_{SW}(H)$ is a Hilbert manifold, modelled on the Hilbert space ${\cal J}_2(H_+,H_-)$, of all Hilbert-Schmidt operators from $H_+$ to $H_-$.
A general element of the group of restricted automorphisms $\operatorname{GL}_{\infty}(H)$, has the block matrix form
$$
\left(\begin{array}{cc}A & B \\ C & D\end{array}\right)\qquad\text{where}\quad \begin{array}{cc} A: H_+\rightarrow H_+, & B:H_-\rightarrow H_+, \\ C:H_+\rightarrow H_-, & D: H_-\rightarrow H_-,\end{array}
$$
with $B,C$ being Hilbert-Schmidt and $A,D$ Fredholm operators. 

In our case the Hilbert space is given by $H:=L^2(S^1,\C)$. Let $H_+$ be the subspace consisting of functions of the form $\sum_{k\leq 0} d_k e^{ikt}$, and $H_-$ the subspace of functions of the form $\sum_{k>0} c_k e^{ikt}$. In fact $H_-\cup\{1\}$, can be identified with the Hardy space $H^2(\overline{\D})$. More generally, we could take as the Hilbert space, all formal Laurent series around the origin, and split them into their regular and principal (singular) parts. 

A related possibility would be to consider only those functions on $S^1$ which have zero mean, or equivalently all exact 1-forms. In the language of Laurent series this gives a splitting into the principal part and the regular part without the constant term. This approach leads to the infinite, generalised Siegel disc, and permits to give a generalisation of the classical period mapping~\cite{KY, NS, N}. 

\subsection{Yur'ev-Krichever embedding into the infinite Siegel disc and the UGM }
It was shown by D.~Yur'ev~\cite{KY}, that the Grunsky matrix provides an embedding (period mapping) of ${\cal M}\subset\Aut({\cal O})$, into the universal Grassmannian.

Let us take the complexified Sobolev space $H:=H_{\C}:=H^{-1/2}(S^1,\C)/\C$, consisting of (generalised) functions with zero mean, and let us define a symplectic form on $H_{\C}$:
$$
\omega(f,g):=\frac{1}{2\pi}\oint_{S^1} f dg~.
$$
With respect to the above bilinear, alternating form, the Hilbert space $H_{\C}$ has a canonical decomposition into two closed isotropic subspaces, as already previously introduced.

The {\bf infinite Siegel disc} ${\cal D}_{\infty}$ consists of operators $Z\in M(H_-, H_+)$, which are symmetric and Hilbert-Schmidt, i.e., they satisfy
\begin{equation}
\label{SiegelDisc}
Z^{\top}=Z,~\qquad\text{and}\quad~1-\overline{Z}^{\top}Z>0~.
\end{equation}
The Kähler potential on ${\cal D}_{\infty}$ can therefore be defined as:~ $-\tr\log(1-Z^*Z)$.

Let $f\in{\cal M}\subset\Aut({\OO})$. Then one can holomorphically associate to $f$ the subspace $W_f\in\Gr(H)$, expressed in terms of the Faber polynomials $F_n$, or equivalently, with the Grunsky matrix $(b_{-n,-m})$, as the space spanned by $1$, and all vectors of the form:
\begin{equation}
\label{YK-map}
w_n(z):=F_n(f(z))=z^{-n}+n\sum_{m=1}^{\infty}b(f)_{-n,-m}\, z^{m}~\qquad n\geq1~.
\end{equation}
In the literature~\cite{ADKP}, a seemingly different mapping was used, but which in fact reduces in the analytic case to the embedding via the Grunsky matrix. Let $H, H_+$ and $H_-$, be as already introduced.  

Then one can associate to a given function $f(z,w)$, which is holomorphic in a neighbourhood of $(0,0)$, the linear operator $T_f$, defined by
\begin{eqnarray}
\label{ADKP10}
T_f:H_+ & \rightarrow & H_-, \\\nonumber
h(z^{-1}) & \mapsto & \res_{w=0}\left( f(z,w) h(w^{-1})\right). 
\end{eqnarray}
Analogously, given a function $g(z^{-1}, w^{-1})$, holomorphic on $(\hat{\C}\setminus\{0\}\times \hat{\C}\setminus\{0\})$ and vanishing at $(\infty, \infty)$, one can associate the operator
\begin{eqnarray}
T_g:H_- & \rightarrow & H_+, \\\nonumber
h(z) & \mapsto & \res_{w=0}\left( g(z^{-1},w^{-1}) h(w)\right). 
\end{eqnarray}
Similar constructions apply for the diagonal entries of the $2\times 2$ block matrix (as just previously introduced). 
Then these operators are continuous and the space $M(H_+, H_-)$, consists of operators of the form $T_f$, with the analogous statement for $M(H_-, H_1)$ and $T_g$, (cf. ~\cite{ADKP} and the reference in there). 

But, this gives an equivalent description to~(\ref{YK-map}), as one can see by applying the definition of the operator to meromorphic functions of the form $z^{n}$,  $n\in\Z$.
\subsection{Regularised Determinants and Determinant Lines}
In String Theory but also in Conformal Field Theory, the partition function is considered to be a section of a determinant line bundle. Here we shall briefly recall how one can define them in different ways. The necessity to use the partition function comes from the fact, that it is the canonical observable for a domain on which one intends to derive a conformally invariant family of measures on simple paths. Physically, this may involve an underlying lattice model with appropriate boundary conditions. Mathematically, it translates into a tensor product of the bulk contribution(s) with the boundary contribution(s), which are forms of a  specific weight. This is the basic idea behind the approach to SLE and its generalisations, as originally introduced in~\cite{BF_corr, FK, F, K, KS}.  

Let us start by explaining, how one can associate to a univalent map on the unit disc a determinant respectively a determinant line. There are at least five relevant ways of doing so. More details concerning this section can be found in~\cite{F1, OPS, PS, ADKP, TT}.

To every Jordan domain $D$, we can associate the determinant of the Laplacian $\Delta$ (with respect to a fixed reference metric, e.g., the Euclidean metric, and with Dirichlet boundary conditions), i.e.,  
$$
\det(\Delta_D):=\det(\Delta_{g_{\operatorname{Eucl.}}}):=\prod_{\text{eigenvalues}}\lambda_n:=e^{-\zeta'(0)}~.
$$
So we get a trivial line bundle over ${\cal M}\subset\Aut({\cal O})$, by associating to a point $f$, the value of the determinant $\det(\Delta_f)$, in the corresponding domain $D_f:=f(\D)$,  i.e.,
\[
\begin{CD}
\det(\Delta_f)\\
@V \pi VV  \\
{\cal M}\subset\Aut({\cal O})
\end{CD}
\]
The uniformising map provides us also with a natural connection which allows us to compare the regularised determinants at different points. It has its origin in Polyakov's string theory~\cite{P}.   
Let us consider the space ${\cal F}$, of all flat metrics on $\D$ which are conformal to the Euclidean metric, obtained by pull-back. Namely, for a Jordan domain $D$, let $f:\D\rightarrow D$ be a conformal equivalence, and define
$$
\phi:=\log|f'|~.
$$
This gives a correspondence of harmonic functions on the unit disc with the category of simply connected domains, and by Weyl rescaling, with ${\cal F}$ itself, via
$$
ds=|f'||dz|=e^{\phi}|dz|~.
$$
To fix the $\operatorname{SU}(1,1)$-freedom, which gives classes of isometric metrics, we have to work with equivalence classes. So, e.g. $0$ corresponds to the orbit of the Euclidean metric under $\operatorname{SU}(1,1)$.

The connection is given by the {\bf Polyakov-Alvarez} formula (cf. e.g.~\cite{OPS}):
\begin{equation}
\label{PA_rel}
\det(\Delta_f)=e^{-\frac{1}{6\pi}\oint_{S^1}(\frac{1}{2}\phi{*}d\phi
+\phi|dz|)}\cdot\det(\Delta_{\D})~.
\end{equation}

Then the partition functions of the corresponding domains (for a two-dimensional target space and without the boundary contributions) are related by: 
\begin{equation}
\label{PFunction_rel}
Z_f=e^{\frac{1}{6\pi}\oint_{S^1}(\frac{1}{2}\phi{*}d\phi
+\phi|dz|)}\cdot Z_{\D}~.
\end{equation}
It is a classical fact (cf. e.g.~\cite{BPZ, KNTY, TT} and references therein), that the infinitesimal variation, induced by a vector field $v$, on a Riemann surface, of a determinant, $\det$, coming from the Green's function, is related to the Schwarzian derivative:
\begin{equation}
\label{Schwarzian}
S(f,z):=\{f;z\}:=\frac{f'''(z)}{f'(z)}-\frac{3}{2}\left(\frac{f''(z)}{f'(z)}\right)^2~.
\end{equation}
In Physics, the variation of $\log(\det)$, is then said to be given by the ``expectation value of the energy-momentum tensor", i.e.,
\begin{equation}
\label{EMT-variation}
\delta\log(\det)=\frac{1}{2\pi i}\langle T(v)\rangle~,
\end{equation}
which, as is well known, has the transformation properties of a projective connection. 
Let us note, that this transformation property is at the origin of the so-called ``restriction martingale" from SLE, which was found in~\cite{LSW}, and then explained in~\cite{FK}.

The determinant line bundle over the infinite Grassmannian is a holomorphic line bundle $\Det$. The fibre $\Det(W)$ at $W\in\Gr(H)$, can be seen to be the ``top exterior power" of $W$, and which can be constructed as follows~\cite{PS, K}:  for a so-called admissible basis $w=\{w_k\}$ of $W$ and $\lambda\in\C$ it is the formal semi-infinite wedge product
$$
\lambda w_{-d}\wedge w_{-d+1}\wedge_{-d+2}\wedge\cdots~.
$$
If we change the basis $w\mapsto w'$ with $w={\bf t} w'$, then $\lambda\mapsto\lambda\det({\bf t})$. The union of all such one-dimensional complex vector spaces $\Det(W)$ for $W\in\Gr(H)$, is the line bundle $\Det_{\Gr}$. Differently, for $W$ let
\begin{equation}
\label{det_ADKP}
\Det(W):=\bigwedge^{\text{max}}\left(\ker(\pr_+|_W)\right)^*\otimes\bigwedge^{\text{max}}\coker(\pr_+|_W)~.
\end{equation}
The sections of $\Det^*$, the dual of the determinant line bundle, constitute the {\bf free Fermionic Fock} space, $\Gamma(\Det^*)$.

Let us summarise the various relations between the spaces of univalent functions and the infinite Grassmannian in the following diagram: 
\[
\begin{CD}
\label{trans_action}
@.\det(\Delta_f)@>\text{up to a}>\text{factor}>\operatorname{Det}\\
@.@VVV@VVV\\
\Diff_+(S^1)/S^1\cong S_0@>\text{dense}>>S\subset\Aut_+({\cal O})@>\text{Yur'ev}>\text{Krichever}>\Gr(H)
\end{CD}
\]

\section{Virasoro modules}
\label{Vir_module}
We are going to recall two representations of the Virasoro algebra. The first one could be called ``universal" in our context, and the second one, is a particular realisation of it. Namely, it gives the ``universal deformation of the complex disc". 
\subsection{Virasoro action on the Universal Grassmannian}
The complex vector fields~(\ref{Vir_Top_base}), $\{e_n\}_{n\in\Z}$, on the circle, which satisfy the commutation relations of the Witt algebra, have a representation~(\ref{LieAction}) in terms of the Lie fields ${\cal L}_{e_n}$, and as we know act transitively on $\aut(\OO)$, i.e., the Lie algebra homomorphism
$$
\{e_n\}\rightarrow\{{\cal L}_{e_n}\}=\operatorname{Vect}\left(\Aut({\OO})\right)
$$
has the property that for every $f\in\Aut({\OO})$ the evaluation map
$$
\rho_f: \{e_n\}_{n\in\Z}\rightarrow T_f \Aut({\OO})\simeq\der_0(\OO),
$$ is surjective.

Analogously, there exists  a representation of the $\{e_n\}$ in terms of  infinite matrices for the Grassmannian~\cite{K1, PS}. 

Namely, the Lie algebra of the Lie group $\operatorname{GL}_{\infty}$, is $\mathfrak{gl}_{\infty}$. It is contained in the bigger Lie algebra $\mathfrak{a}_{\infty}$, which can be described as the infinite matrices, with a finite number of nonzero diagonals. This algebra has a projective representation, and its central extension $\hat{\mathfrak{a}}_{\infty}:=\mathfrak{a}_{\infty}\oplus\C\mathfrak{c}$, has a linear representation.  

The two-cocycle on $\mathfrak{a}_{\infty}$, is given for two elements $a_1, a_2$, by $\tr(c_1\cdot b_2-c_2\cdot b_1)$, where the $b_i, c_j$ are the corresponding block matrices in the decomposition of the $a_k$. 

The group $\widehat{\operatorname{GL}}_{\infty}$, is the central extension of $\operatorname{GL}_{\infty}$. It acts on the determinant line bundle $\Det$, and it covers the transitive action of $\operatorname{GL}_{\infty}$.    

Finally, the orbit of the {\bf Bosonic vacuum}, $1$, under the action of $\operatorname{GL}_{\infty}$, is an infinite-dimensional manifold. Its description is given by {\bf Hirota's bilinear equation}. This space contains, as we shall see, the traces of SLE. So, the connection with the {\bf $\tau$-function} and {\bf Integrable Systems}, is immediate. 

Let us summarise all this:
\noindent
\[
\begin{CD}
\label{trans_action}
\vir_{\C}@>>>\hat{\mathfrak{a}}_{\infty}@>>>\widehat{\operatorname{GL}}_{\infty}@>>>\Det\\
@.@A\text{central}AA@AAA\\
@.\mathfrak{a}_{\infty}@.@.\\
@.@AAA@AAA\\\witt@>>>\mathfrak{gl}_{\infty}@>\exp>>\operatorname{GL}_{\infty}@>\text{transitive}>>\Gr(H)
\end{CD}
\]
\subsection{The Kirillov-Yur'ev-Neretin representation of the Virasoro algebra} 
\label{KYN_Vir_action}
Kirillov-Yur'ev and Neretin gave an explicit description of the action~\cite{KY, N} in terms of the infinite-affine co-ordinates (cf. also~\cite{F1, F})

On the infinite complex manifold $\aut_+(\OO)$ there exists an analytic line bundle $E_{c,h}$ which carries a transitive action of the Virasoro algebra. The line bundle $E_{c,h}$ is in fact trivial, with total space $E_{c,h}={\aut_+(\OO)}\times\C$. We can parametrise it by pairs $(f,\lambda)$, where $f$ is a univalent function and $\lambda\in\C$.  

It carries the following action defined by Y.~Neretin~\cite{N},
\begin{equation}
\label{KY2}
L_{v+\tau{\mathfrak c}}(f,\lambda)=({\cal L}_v f, \lambda\cdot\Psi(f,v+\tau{\mathfrak c}))~,
\end{equation}
where 
\begin{equation}
\label{Neretin}
\Psi_{c,h}(f,v+\tau{\mathfrak c}):=h\oint\left[\frac{wf'(w)}{f(w)}\right]^2v(w)\frac{dw}{w}+\frac{c}{12}\oint w^2 S(f,w)\frac{dw}{w}+i\tau c~,
\end{equation}
and with $S(f,w)$ denoting the Schwarzian derivative~(\ref{Schwarzian}), of $f$. The central element $\mathfrak{c}$ acts fibre-wise linearly by multiplication with $ic$. The appearance of the Schwarzian is not so surprising if one remembers the infinitesimal variation of the regularised determinant. However, in~\cite{KY}, this term was not present. 

One can endow the space of holomorphic sections $|s\rangle\in{\cal O}(E_{c,h})\equiv\Gamma({\aut_+(\OO)}, E_{c,h})$ of the line bundle $E_{c,h}$ with a $\vir_{\C}$-module structure. 
  
Namely, let ${\cal P}$ be the set of (co-ordinate dependent) polynomials on $\cal M$, defined by 
$$
P(c_1,\dots, c_N):\Aut({\OO})/\mathfrak{m}^{N+1}\rightarrow \C~,
$$
with $\mathfrak{m}$ the unique maximal ideal. In the  classical literature on complex variables the above quotient space is called the {\bf coefficient body}. ${\cal P}$ corresponds to the sections ${\cal O}({\cal M})$ of the structure sheaf ${\OO}_{\cal M}$ of ${\cal M}$, and it 
carries an action of the representation of the Witt algebra in terms of the Lie fields ${\cal L}_{n}\equiv{\cal L}_{e_n}$.

Now, ${\OO}(E_{c,h})$ can be identified with the polynomial sections of the structure sheaf, i.e. with ${\OO}({\cal M})$, by choice of a free generator, and the previous trivialisation.  

The action of $\vir_{\C}$ on sections of $E_{c,h}$ can then be written in co-ordinates, with the notation $\partial_n\equiv\frac{\partial}{\partial c_n}$, according to formulae~(\ref{KY2}, \ref{Neretin}) as~(cf.~\cite{KY}),
\begin{eqnarray}
\label{Vir_coord}
\nonumber
L_n & = & \partial_n+\sum_{k=1}^{\infty}c_k\, \partial_{k+n},\quad n>0 \\\nonumber
L_0 & = & h+\sum_{k=1}^{\infty} k\, c_k\partial_k~,\\ \nonumber
L_{-1} &=& \sum_{k=1}^{\infty}\left((k+2)c_{k+1}-2c_1c_k\right)\partial_k+2hc_1~,\\
L_{-2} &=& \sum_{k=1}^{\infty}\left((k+3)c_{k+2}-(4c_2-c_1^2)c_k-a_k\right)\partial_k+h(4c_2-c_1^2)+\frac{c}{2}(c_2-c_1^2)~,
\end{eqnarray}
where the $a_k$ are the Laurent coefficients of $1/f$, $c$ is the central charge and $h$ the highest-weight. 

The conformal invariance of the measures we are going to consider, will naturally lead to highest-weight modules.\\  Now a polynomial $P(c_1,\dots, c_N)\in{\OO}(E_{c,h})$ is a singular vector for $\{L_n\}$, $n\geq1$, if
$$
\left(\partial_k+\sum_{k\geq1}(k+1)c_k\partial_{k+n}\right)P(c_1,\dots, c_N)=0~.
$$
Then, the highest-weight vector is the constant polynomial $1$, and satisfies $L_0.1=h\cdot 1$, $\mathfrak{c}.1=c\cdot 1$. 

The dual ${\cal O}^*(E_{c,h})$, i.e. the space of linear functionals $\langle s|$ on ${\cal O}(E_{c,h})$, can again be identified with ${\cal O}({\cal M})$ via the pairing,
$$
\langle P,Q\rangle:=P(\partial_1,\dots,\partial_N)Q(c_1,\dots,c_N)|_{c_1=\dots=c_N=0}~,
$$
and the action of $\vir$ can then similarly be given explicitly in co-ordinates as in equations~(\ref{Vir_coord}), with the roles of $c_k$ and $\partial_k$ interchanged.

Analogously, the dual space ${\OO}^*(E_{c,h})$ can be endowed with a $\vir_{\C}$ action. The singular vector for the corresponding (irreducible) Verma module $V_{c,h}$ with respect to $\{L_{-n}\}$, $n\geq1$, is again the constant polynomial $1$.

We have summarised the geometry and the various actions in the following commutative diagram of short exact sequences
\[
\begin{CD}
\label{trans_action}
0@>>> \C@>>>\vir_{\C}@>>>\operatorname{Witt}@>>>0\\
@.  @VVV@VVV@VVV@.\\
0 @>>> \C@>>>\Theta_{E_{c,h}}@>>>\Theta_{\cal M}@>>>0\\
@.@VVV@VVV@VVV@.\\
0@>>> \C@>>>E_{c,h}@>>>{\cal M}@>>>0
\end{CD}
\]
where $\Theta_{X}$ denotes the respective tangent sheaf. 

\section{The SLE path space}
Let us relate now the general theory, as developed in the previous sections, to SLE. 

\subsection{The SLE path space on the Grassmannian} 
Let us consider the set ${\cal W}_{0,S^1}(\D)$, of all Jordan arcs on the closed unit disc $\overline{\D}$, such that one endpoint is the origin $0$, the other endpoint is on the boundary $S^1$, and all other points are in the interior, $\D$.

If we fix a parametrisation $\gamma(t)$, of such an arc, starting at the boundary, i.e., $\gamma(0)\in S^1$, then we can obtain a family of ``standard" normalised univalent maps $g_t$,  such that the arc up to time $t$, $\gamma[0,t]$, satisfies $\gamma[0,t]=\overline{\D\setminus g_t(\D)}$. The family $\{g(t,z)\}$, which forms a subordination chain, is then of the form
$$
g(t,z)=e^{\lambda(t)}z+a_2(t)z^2+\cdots~\in\Aut({\OO}),
$$ 
with respect to its power series development, where $\lambda(t)$ is a strictly monotonic real function.
Then we have
\begin{prop}
Up to reparametrisation,  the set ${\cal W}_{0,S^1}(\D)$, maps injectively (and continuously) into the space of simple paths in ${\cal M}\subset\Aut({\OO})$, starting at the identity $z$. Equally, it can be mapped one-to-one into the space of simple paths in the Grassmannian $\Gr(H)$, starting at $H_+$.
\end{prop}  
Injectivity follows from the fact that  
the traces are compact subsets of the domain, which is a Hausdorff space. If two traces, with the exception of the endpoints are different, there are at least two interior points of the curves which are different. Hence, we can choose open disjoint neighbourhoods, around these two points.  The family of mappings for one of the curves at some instance maps the point onto the boundary, for the other never. Then, it follows from the Open Mapping theorem and the Identity theorem for analytic functions that the corresponding traces in ${\cal M}$, must have points which are not the same, which in turns shows injectivity. 
\subsection{Hypo-Ellipticity and sub-Riemannian Geometry}
The SLE equation~(\ref{Loewner-Ito}) can be expanded into a Laurent series around infinity. Namely, the hydrodynamically  normalised functions $f_t$, satisfy:
$$
f_t(z)=z-\sqrt{\kappa} B_t+\frac{2t}{z}+\cdots\quad\in\Aut_+({\cal O}_{\infty})~,
$$
where the coefficient $b_0=-\sqrt{\kappa} B_t$, is given by the Brownian motion. To Taylor expand $1/f_t$, we can define recursively (time-dependent) polynomials $p_k=p_k(t)$, in the coefficients $b_i=b_i(t)$. Then $p_1\equiv1$ and $p_k=-\sum_{i=0}^{k-1} b_i p_{k-1-i}$ for $k\geq2$, such that $f_t(z)^{-1}=\sum_{n=1}^{\infty} p_n z^{-n}$. So the SLE equation translates into the hierarchy (projective limit) of SDEs:
\begin{eqnarray*}
db_0(t) & = & -\sqrt{\kappa}dB_t \\
db_1(t) & = & 2 p_1(b_0)dt\\
&\dots&\\
db_n(t) & = & 2 p_n(b_0,\dots, b_{n-1}) dt\\
&\dots&
\end{eqnarray*}

On the coefficient body ${\cal M}_{N}$, $N\geq0$, we have as finite dimensional approximations for the drift vector $\underline{b}_N$ and the $(N+1)\times 1$ diffusion coefficient $\sigma_N$:  
$$
\underline{b}_N=2\cdot\left(\begin{array}{c}0 \\1 \\ p_2 \\\vdots \\p_N\end{array}\right)~,\qquad~\sigma_N =\left(\begin{array}{c}-\sqrt{\kappa}  \\0 \\ 0 \\\vdots  \\0 \end{array}\right)~.
$$
Now, by taking the projective limit we obtain the generator $A_{\infty}$, of the flow on ${\cal M}\subset\aut_+({\OO}_{\infty})$, corresponding to the L{\oe}wner equation (\ref{Loewner-Ito}) for some fixed  $\kappa$,~\cite{BB} (and related in~\cite{FW1, FW2}):
\begin{equation}
\label{L_generator}
A_{\infty}=\varprojlim\left(\frac{\kappa}{2}\frac{\partial^2}{\partial b_0^2}+2\sum_{k=1}^N p_k(b_0,\dots, b_{k-1})\frac{\partial}{\partial b_k}\right)~,
\end{equation}
which is driven by one-dimensional standard Brownian motion of variance $\kappa$, and where the polynomials $p_k(\underline{b})$ in the drift vector, are defined on the coefficient body ${\cal M}_N$.

The generator $A_{\infty}$, of the diffusion process can be written in Hörmander form, in terms of the tangent vector fields in the affine co-ordinates $\{b_k\}$, by applying the notational conventions for operators acting on polynomial sections of the structure sheaf as 
\begin{equation}
\label{Witt_SLE_gen}
A_{\infty}=\frac{\kappa}{2}(L_{-1}^{\infty})^2-2L_{-2}^{\infty}~,
\end{equation}
where
$$
L_{-1}^{\infty}:=-\frac{\partial}{\partial b_0},\qquad\text{and}\qquad L^{\infty}_{-2}:=-\sum_{k=1}^{\infty}p_k(\underline{b})\frac{\partial}{\partial b_k}~.
$$
But as $L_{-1}^{\infty}$ and $L_{-2}^{\infty}$, satisfy the commutation relations of the Witt algebra, they span the whole tangent space and since we know that this Lie algebra acts transitively, the strong Hörmander condition is satisfied.  Therefore, the resulting flow is hypo-elliptic~\cite{K,F}, and the corresponding geometry sub-Riemannian.

Let us note, that the above discussion clearly links SLE with ``Rough Path Theory" (cf.~\cite{F09}).

\section{Statistical Mechanics and Virasoro action on measures}
In the previous sections we have seen that the conformally invariant measures on simple paths are characterised by the diffusion constant $\kappa$. 

Given a simply connected planar domain $D$, conformally equivalent to the unit disc, and two distinct boundary points $A$ and $B$, then one can consider ${\cal W}={\cal W}(D_{A,B})$, the set of Jordan arcs in $D$, connecting $A$ with $B$. A particular trace $\gamma\in{\cal W}$, corresponds then to an elementary event. However, the $\sigma$-algebra ${\cal F}_{D_{A,B}}$, we are going to use, contains also all Jordan arcs, starting either at $A$ or $B$, without connecting the other marked boundary point. However, the set of all these events will have measure zero, i.e. the measure is supported on ${\cal W}$. 

The set $\{\mu_{D_{A,B}}\}$ of all measures on the traces, forms a convex cone. The subset of all probability measures ${\cal P}_{D_{A,B}}$, called the {\bf state space}, is a convex set, i.e. with $\mu_1, \mu_2$ also any convex combination $\mu = \alpha_1 \mu_1 + \alpha_2 \mu_2$, $\alpha_{1,2}\geq0$ and $\alpha_1 + \alpha_2 = 1$ is a (mixed) state. The points in the closure of the state space are the Dirac measures $\delta_{\gamma}$.

If we assume that on our domain a lattice model from statistical mechanics is defined, then the thermodynamical equilibrium at the absolute temperature $T > 0$ is described by the Gibbs state 
$$
d\mu_G(\sigma) :=\frac{1}{Z} e^{-\beta E[\sigma]},\qquad\beta :=\frac{1}{T}\qquad\text{(inverse temperature)}
$$
where the numerical pre-factor $Z > 0$, gives the normalisation, such that the sum of all elementary events ads up to $1$. It is defined as 
$$
Z :=\sum_{\{\sigma\}} e^{-\beta E[\sigma]},
$$
where $E[\sigma]$ is the energy of the configuration $\sigma$. $Z$ is the (discrete) partition function. By choice of appropriate boundary conditions (bc), every configuration compatible with the bc, creates an element in ${\cal W}$. The Gibbs measure corresponds then, at the discrete level, to the canonical state, and in the continuum, it is given by the exponential of a Lagrangian density.  

\begin{figure}[htbp!]
\begin{center}
\includegraphics[scale=0.5]{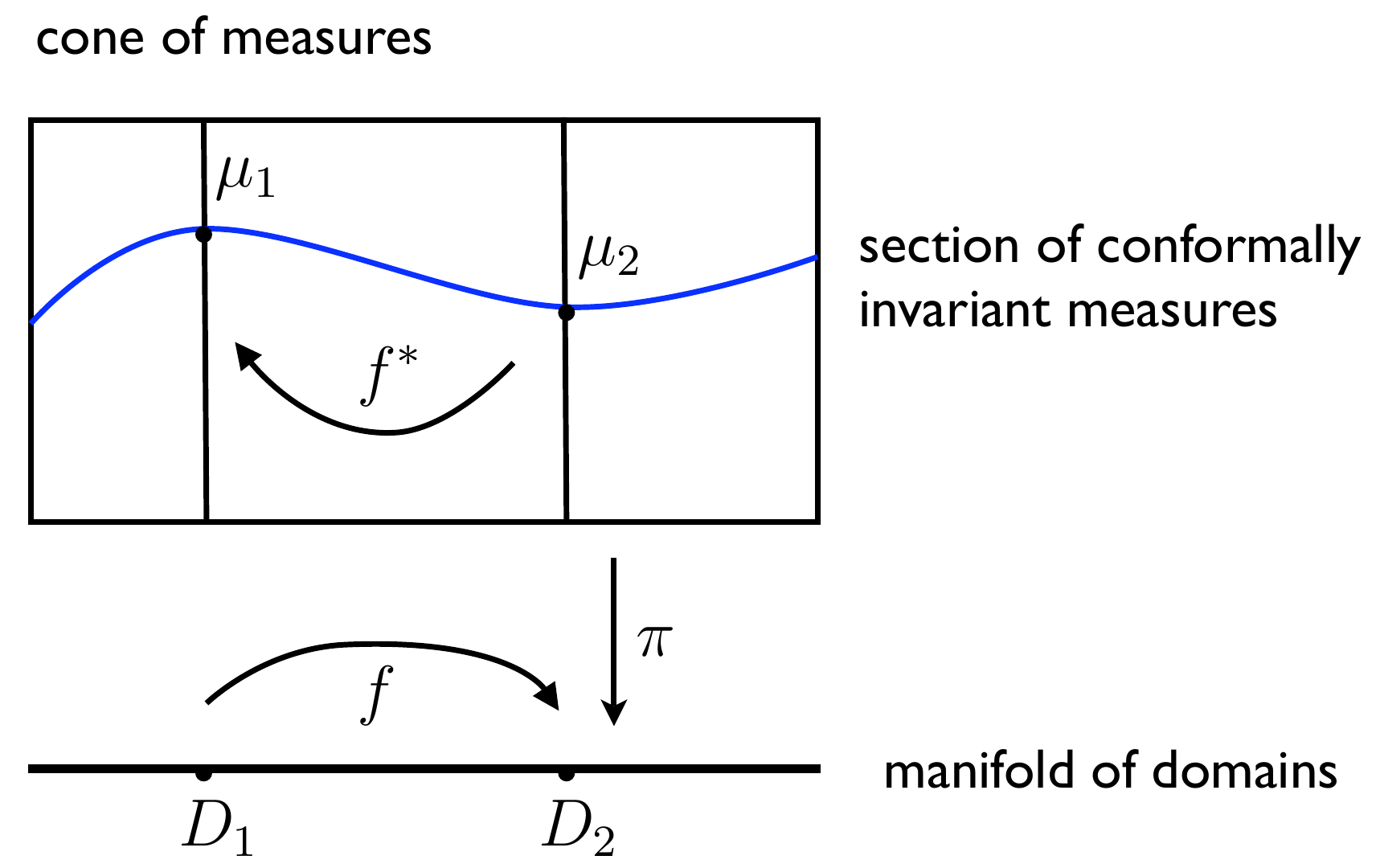}
\caption{The fibre bundle of measures over the manifold of (slit) Jordan domains.} 
\label{states}
\end{center}
\end{figure}

If we focus on the measures which can be obtained as the continuum limit of lattice curves, then the following two properties are found to be natural to hold for these measures:
\begin{description}
  \item[Markovian property] If $\gamma$ is a random Jordan arc, and $\gamma'\subset\gamma$ a sub-arc, with $A$ as common starting point, and with $A'$ as endpoint, then the conditional distribution of $\gamma$ given $\gamma'$ is 
\begin{equation}\label{E:markov}
\mu_{D_{A,B}|{\gamma'}}=\mu_{(D\backslash\gamma')_{A',B}}~.
\end{equation} 
\end{description}
Note, that this property is expected to be true for the scaling limit of several models, even off the critical point.

The second property is the so-called ``conformal invariance", which, however, should be valid only at the critical point. It was conjectured and investigated by {\bf M.~Aizenman, R.~Langlands, P.~Pouilot, Y.~Saint-Aubin}~\cite{Ai, LPS}, and then established by O.~Schramm~\cite{Sch}, for the loop-erased random walk and the uniform spanning tree. 
\begin{description}
  \item[Conformal invariance (ALPS-functor)] 
Let $f:D\rightarrow D'$, be a conformal equivalence, so that the boundary points $(A,B)$ of $D$ are mapped onto the points $(A',B')$ on the boundary of $D'$, preserving the order. Then the image measure $f_*\mu_{D_{A,B}}$ is the same as $\mu_{D'_{A',B'}}$, the measure which is indigenously obtained as the continuum limit of lattice curves from $A'$ to $B'$ in $D'$.
\end{description}
The above categorical statement can be represented by the following commutative diagram
\[
\begin{CD}
D_{A,B}@>f>> D'_{A',B'}\\
@V F_{\text{ALPS}} VV  @V F_{\text{ALPS}}VV\\
\mu_{D_{A,B}} @>f_*>> \mu_{D'_{A',B'}}
\end{CD}
\]

The category $\text{\texttt{JDom}}_{\bullet,\bullet}$ of (generalised) Jordan domains with two marked and ordered boundary points, can be parametrised by an infinite manifold. Namely, if we consider first the set 
$\text{\texttt{J}}^{\infty}$ of smooth Jordan curves,  then one has the double quotient~\cite{AMT}
$$
\text{SU}(1,1){\setminus}\Diff_+(S^1) / \text{SU}(1, 1)
$$
as the base space, and as fibre model of the total space, the torus  (minus the diagonal), i.e., $S^1\times S^1\setminus\{\text{diagonal}\}$.
The sections correspond then to domains with two distinct, marked boundary points.

As conformally invariant measures are translational invariant, it is enough to look at $\text{\texttt{J}}^{\infty}_0$, i.e., all smooth Jordan curves surrounding the origin. But we already know, that this set is a submanifold of $\Aut({\OO})$, which also contains slit Jordan domains. 
Therefore we have 
\[
\begin{CD}
\{\text{measures on Jordan arcs}\}@>>>{\cal M}\times (S^1\times S^1\setminus\{\text{diagonal}\})@>\pi>>{\cal M}\hookrightarrow\Aut({\OO})~.
\end{CD}
\]
Then the trivial bundle, with fibre (almost) a torus, parametrises, up to one real positive constant $\kappa$, conformally invariant probability measures on simple paths, which connect two distinct boundary points. 

Instead of the above bundle of measures on Jordan arcs, we have also to regard line bundles $Z_{c,h}$, which represent the partition function of a model from statistical mechanics, defined on domains with given boundary conditions. The situation is similar to the already introduced bundles $E_{c,h}$.
\subsection{Deformations of conformal measures and the partition function valued martingale}
We have introduced the bundle of measures, with convex sets as fibres. A subset of this space, namely the conformally invariant measures on bordered domains, obtained from lattice models in the scaling limit at criticality, depend on two parameters $(c,h)\in\R^2$,~(cf.~\cite{F1}). 

Now, to make the link between the SLE description and the Conformal Field Theory descritpion, we shall deform the state by conditioning it on an appropriately growing sub-arc,~(cf.~Fig.~\ref{states}). By the Markovian type property this is the same as considering a family of randomly evolving slit domains. 

\begin{figure}[htbp!]
\begin{center}
\includegraphics[scale=0.5]{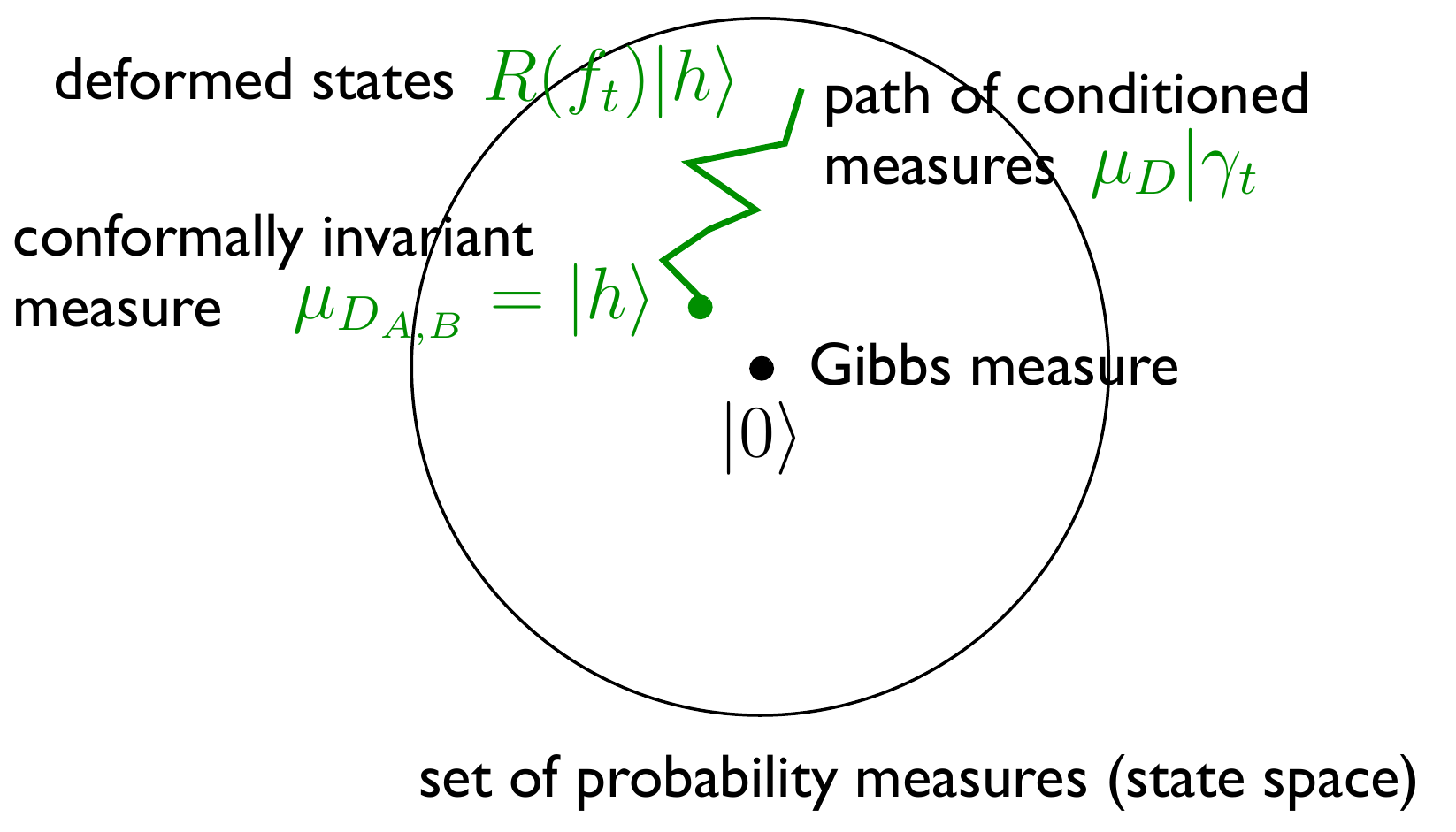}
\caption{The convex set of probability measures ${\cal P}_{D_{A,B}}$, on a fixed domain $D_{A,B}$. The Gibbs state $|0\rangle$, the conformally invariant state $\mu_{D_{A,B}}=:|h\rangle$, and a Virasoro deformed state $R(f_t)|h\rangle$, which is equal to the conditional measure $\mu_D|\gamma_t$.} 
\label{states}
\end{center}
\end{figure}

Let us fix our reference domain $D_{A,B}$,  and a conformally invariant measure $\mu_D\equiv\mu_{D_{A,B}}$. This measure defines a state $|h\rangle\in{\cal P}_{D_{A,B}}$. Consider a random Jordan arc $\gamma([0,t])$, which by abuse of notation we shall denote $\gamma_t$, connecting the  boundary point $A=\gamma(0)$ with the interior point $A'=\gamma(t)\in D$. Further, let $\{\gamma_t\}$, be the measurable set of traces $\gamma'$, which contain $\gamma_t$ as a sub-arc, i.e. $\gamma_t\subset\gamma'$. Then, the conditional probability
$$
\mu_D|\gamma_t(\bullet):=\frac{\mu_D(\bullet\cap\{\gamma_t\})}{\mu_D(\{\gamma_t\})},
$$
gives a new measure on the same underlying measurable space.
But, at is was shown in~\cite{FK}, the ratio of the partition function $Z_{\gamma_t}:=Z_{D\setminus\gamma_t}$, of the slit domain and of the original domain $Z_D$,  ``corresponds to the volume of paths containing the arc $\gamma_t$". Therefore,
$$
\mu_D(\{\gamma_t\})=\frac{Z_{\gamma_t}}{Z_D}\leq 1,
$$
and hence
$$
\mu_D|\gamma_t(\bullet)=\underbrace{\frac{Z_D}{Z_{\gamma_t}}}_{=:M_{\gamma_t}}\cdot \mu_D(\bullet\cap\{\gamma_t\}).
$$

Let $f_{\gamma_t}:D\setminus\gamma_t\rightarrow D$, be a conformal equivalence. Then by the Markovian property and the conformal invariance of the measure, we have
$$
\mu_D|\gamma_t=\mu_{D\setminus\gamma_t}=f_{\gamma_t}^*\mu_D.
$$
Note that $f_{\gamma_t}^*\mu_D=(f^{-1}_{\gamma_t})_*\mu_D$, i.e., the pull-back measure is given by the inverse function $f^{-1}_{\gamma_t}:D\rightarrow D\setminus\gamma_t\subset D$.

Let us recall that the elementary events are the random arcs $\gamma$. So, for the Radon-Nikodym derivative we obtain, 
\begin{equation}
\label{RN-martingale}
\frac{d\mu_D|\gamma_t}{d\mu_D}=M_{\gamma_t}=\frac{Z_D}{Z_{\gamma_t}}.
\end{equation}
Analogously to the reasoning which led to the derivation of the nature of the driving function of SLE, we can conclude that 
$M_{\gamma_t}\equiv M(\gamma,t)=M_t(\gamma)$, has to be a (local) martingale on the path space, if it comes from a conformally invariant measure on paths. 

So, what we have gained is, that equation~(\ref{RN-martingale}) and the martingale property give the explicit relation between SLE and CFT, as we will show.

Let us summarise the content of the fundamental equation~(\ref{RN-martingale}):
\begin{thm}
The operation of conditioning on a sub-arc, gives a deformation of a conformally invariant measure, on a fixed  measurable space, supported  by Jordan arcs, as obtained from Statistical Mechanics models. The deformation space is a subset of the absolutely continuous measures, with respect to the undisturbed measure, where the Radon-Nikodym derivative is given by the inverse ratio of the partition functions of the underlying domains.
\end{thm}

As next we are going to describe in detail the structure of the martingales $M_{\gamma_t}$, and to show that the deformations are generated by the Virasoro algebra.

The physical partition function $Z$, is mathematically, a tensor product of forms of weight $c$ and $h$, i.e., $Z=Z_{c,h}=Z_c\otimes Z_h$, defined as sections of a line bundle.  Namely, we have 
\begin{thm}[FKK 2003;~\cite{FK, K, F, KS}] Let a configuration $D_{A, B}$, consisting of a simply connected domain $D$, with metric $g$ smooth up to the boundary and two marked boundary points $A, B$, with analytic local co-ordinates, be given. For $(c,h)\in\R^2$, the partition function $Z_{c,h}(D)$ of chordal $\operatorname{SLE}_{\kappa}$ is: 
$$
Z_{c,h}(D) =  |{\det}_D|^{\otimes c}\otimes |T^*_A\partial D|^{\otimes h}\otimes |T^*_B\partial D|^{\otimes h}=\det(\Delta_D)^{-c}\cdot Z_h(\partial D)~, 
$$
where $Z_h(\partial D)$ is the boundary contribution, $c$ the central charge, and $h$ the highest-weight. (The connection between $c,h,\kappa$ shall be revealed later!)
\end{thm}
So, the structure of the above tensor product is composed of the contribution from the bulk, which is expressed by the regularised determinant of weight $c$, and a contribution from a form of weight $h$, restricted to the boundary of the domain. The latter is motivated by the notion of  ``boundary condition changing operators".
In our framework, these operators can be seen as deforming the Gibbs state into a new measure, that has a specific transformation property under conformal maps. These operators act like a multiplication of the integrand with a characteristic function, when calculating the Gibbs sum, respectively the Functional Integral. The particularity is, that these operators can be parametrised by geometrical data, namely the ``location of the insertion on the boundary".

Therefore, the density $M_{\gamma_t}$ depends on the above two parameters $c$ and $h$, i.e., $M_{\gamma_t}=M_{c,h}(\gamma,t)$.

For a moment let us focus on $\tilde{M}_{\gamma_t}$, coming from the regularised determinant part of $M_{\gamma_t}$. By eq.~(\ref{PFunction_rel}) and the semi-group property for regularised determinants~(cf. \cite{F1}), we have for the part without the boundary contribution,  
\begin{equation}
\label{Partitionfunction}
Z_c(t)=\tilde{Z}_{\gamma_t}=e^{\frac{c}{6\pi}\oint_{S^1}(\frac{1}{2}\phi_{\gamma_t}{*}d\phi_{\gamma_t}
+\phi_{\gamma_t}|dz|)} \tilde{Z}_D
\end{equation}
where,  $g_t:=f^{-1}_{\gamma_t}:D\rightarrow D\setminus\gamma_t$, is a biholomorphic map, and $\phi_{\gamma_t}:=\log|\frac{\partial}{\partial z}f^{-1}_{\gamma_t}|$. Now, because of the semi-group property, the determinant can be written as an exponential of infinitesimal conformal maps. We have already seen in~(\ref{EMT-variation}), that the infinitesimal variation of a regularised determinant is given by the Schwarzian derivative. This has been axiomatised  in CFT, as the canonical transformation property of the energy-momentum tensor. 

At this point, let us briefly recall the following facts from the general theory of vertex operator algebras~(cf.~\cite{FbZ}).\\ 
A (local) conformal transformation $\rho$, induces an endomorphism  $R(\rho)$, of the CFT Hilbert space. This corresponds to a representation of the Lie group $\Aut({\cal O})$, on the Hilbert space, by exponentiating the non-negative part of the Virasoro algebra $\Vir_{\geq 0}$. The operator insertions of primary fields $\Xope$, transform homogeneously, whereas the stress-energy operator $T$, changes in-homogeneously, i.e.
\begin{eqnarray}
\Xope (y_i) &=& R(\rho) \Xope \Big(\rho(y_i)\Big)R(\rho)^{-1} ~
\prod_i \Big( \rho'(y_i) \Big)^{h_i}~, \\
T(z) &=& R(\rho)  T\Big(\rho(z)\Big)  R(\rho)^{-1} + \frac{c}{12}
{S\rho}(z)\,\unit~,
\end{eqnarray}
where $S\rho$ is the Schwarzian derivative of $\rho$~(cf. eq.~(\ref{Schwarzian})), and the prime ${}'$ denotes differentiation with respect to the variable $z$.

Now for $D_{A,B}$, let $g_{\gamma_t}:D\rightarrow D\setminus\gamma_t$, be a conformal equivalence such that $g_{\gamma_t}(A)=\gamma(t)$ and $g_{\gamma_t}(B)=B$.

Then following~\cite{FK}, we find for~(\ref{RN-martingale}):
\begin{equation}
\label{FK46}
M_{\gamma_t}\equiv M_{c,h}(\gamma,t)=
|g_{\gamma_t}'(A)|^{h} |g_{\gamma_t}'(B)|^{h}\cdot {\cal R}\,\exp\left(-\frac{c}{6} \int_0^t 
{Sg}_{\gamma_s}(A)\,\d s\right)~,
\end{equation}
where ${\cal R}$ denotes that the exponential expression has to be suitably regularised.  
In particular, if in~Thm.~6.5. in~\cite{LSW}, the hull $A$, is defined as $A:=\{\gamma_t-\epsilon\}\cup\{\gamma_t+\epsilon\}$, and the limit  $\epsilon\to0$, is properly taken, then one recognises~(\ref{FK46}) for $t\to\infty$, as the Radon-Nikodym derivative, of that situation. 

\begin{figure}[htbp!]
\begin{center}
\includegraphics[scale=0.4]{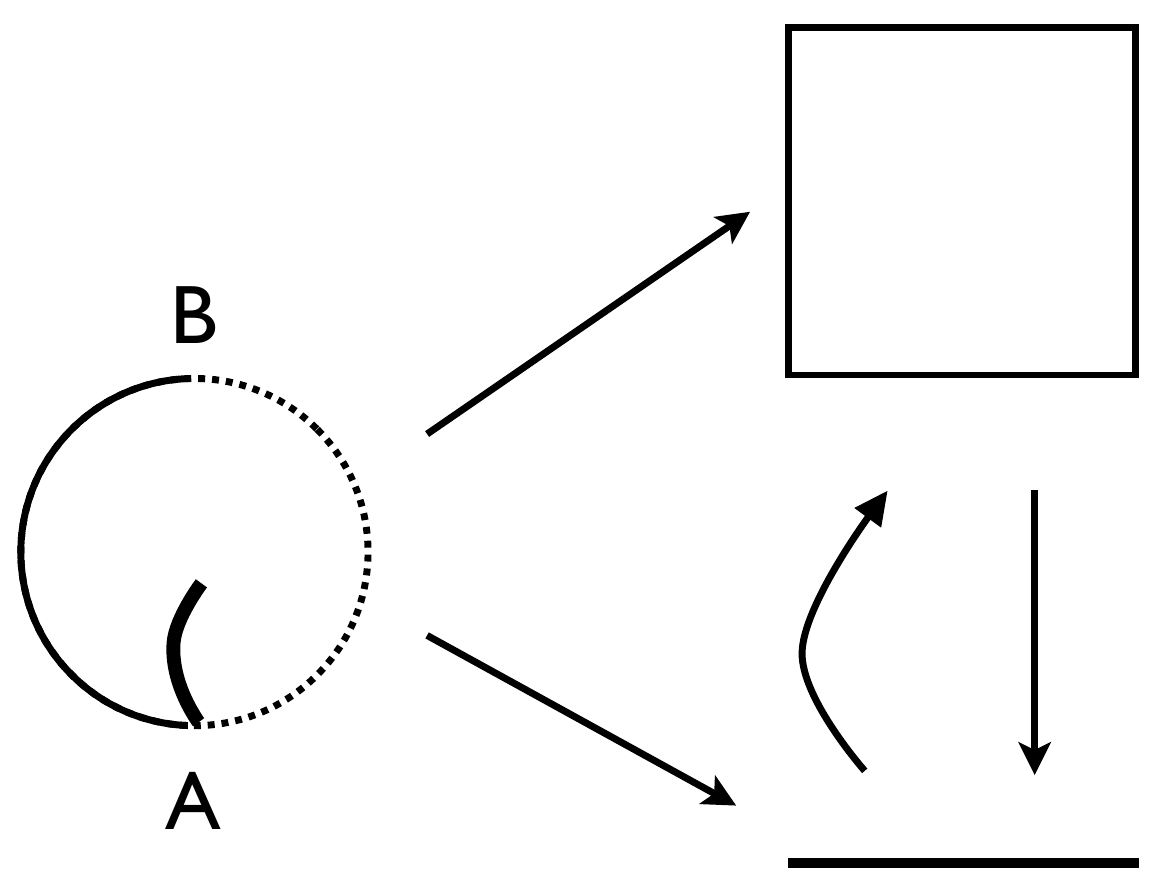}
\caption{Canonical commutative diagram: ``SLE = commutativity".} 
\label{SLEFactor}
\end{center}
\end{figure}

But we already know, as we work with conformally invariant measures on paths, that the projection of the process in the total space onto the base, has to correspond to some $\operatorname{SLE}_\kappa$ in the path space~(cf.~Fig.~\ref{SLEFactor}). This implies, that the family of uniformising maps $g_{\gamma_t}$, is a solution of the related inverse L{\oe}wner chain, and the driving function $U_t$ has to be $\sqrt{\kappa} B_t$ for some $\kappa$.\\
A direct calculation, applying It\^{o}'s formula, as for the ``restriction martingale" in~\cite{LSW}, gives the necessary condition for the expression~(\ref{FK46}) to be a martingale:

\begin{thm}
Given a pair $(c,h)\in\R^2$, and $\kappa>0$, then the density $M_{c,h}(\gamma,t)$ is  a martingale, iff the following relations hold:
\begin{equation}
\label{chkappa}
c=\frac{(6-\kappa)(3\kappa-8)}{2\kappa}\qquad\text{and}\qquad h=\frac{6-\kappa}{2\kappa}~.
\end{equation}
\end{thm}
Instead of projecting the partition function valued martingale $M_t$ down, we could equally well lift the SLE process $f_t(\gamma)$ up to $Z_{c,h}$. But then, as we know, the sections $s:{\cal M}\rightarrow Z_{c,h}$, of the twisted determinant line bundle, have to be such that the resulting process
\begin{equation}
\label{lifted_proc}
Z_{c,h}(t,\gamma):=s\circ f_t(\gamma)~,
\end{equation}
is a martingale. 

As we previously discussed, and because the bundle $Z_{c,h}$ is trivial,  the space of sections $\Gamma({\cal M},Z_{c,h})\equiv{\cal O}({\cal M})$, is given by polynomial functions, once a trivialisation has been chosen, and which are dense in the space of all (real)-analytic sections. Therefore, if the $(a_n)$ are the co-ordinates in ${\cal M}$, the expression~(\ref{lifted_proc}) can be written as (a polynomial approximation)
\begin{equation}
\label{ }
Z_{c,h}(t,\gamma)=s_{\text{polyn.}}(a_1(t,\gamma), a_2(t,\gamma),\dots,a_N(t,\gamma),\dots)~,
\end{equation}
where the notation $a_n(t,\gamma)$ expresses the fact that the coefficients are random variables, i.e. they depend on the curve $\gamma$ and time $t$, and with $s_{\text{polyn.}}$ a polynomial section (function). 

The density $M_{\gamma_t}$ can be written as the ratio of the constant section $s_D\equiv Z_D$ and a section $s_{\gamma}$, such that $Z_{\gamma_t}\in \operatorname{im}(s_{\gamma})$, for all $t\geq0$. 

Now, let us consider what conditions the sections have to satisfy in order that the lifted process is a martingale. 

Therefore, let us apply It{\^{o}}'s formula, in our chosen trivialisation, to (\ref{lifted_proc}). Then, we find:
\begin{equation}
\label{ }
dZ_{c,h}(t)=-dW_t (\hat{L}_{1} s)(f_t)+dt\left((\frac{\kappa}{2}\hat{L}^2_{1}+2\hat{L}_{2})s\right)(f_t)
\end{equation}
where 
\begin{equation}
\label{ }
\hat{L}_{1}:=\frac{\partial}{\partial b_0}\qquad\text{and}\quad~\hat{L}_{2}:=\sum_{k=1}^{\infty} p_k(\underline{b})\frac{\partial}{\partial b_k}~.
\end{equation}
It is important to note, that although the above differential operators $\hat{L}_{1}$ and $\hat{L}_{2}$, look the same as in~(\ref{Witt_SLE_gen}), they are part of the Virasoro algebra with non-trivial central charge, and of the dual representation. Their explicit form is analogous to~(\ref{Vir_coord}), but the representation is taken around infinity. 

\begin{prop}[FKK~\cite{FK,K}, BB~\cite{BB}; 2003]
The lift of an SLE-process, $s\circ f_t$, is a martingale, if the section $s$, is in the kernel of the second-order differential operator $\hat{A}_{\infty}:=\frac{\kappa}{2}\hat{L}_{1}^2+2\hat{L}_{2}$, corresponding to an element from the universal enveloping algebra, in the dual representation.
\end{prop}
As we know from Section~\ref{KYN_Vir_action}, (cf.~\cite{KY}), the Virasoro algebra acts in form of differential operators on the space of sections. It can be endowed with a (graded) Virasoro  module structure.  Similar to the Bosonic Fock space representation of the Virasoro algebra, acting on the constant section $1$, generates a highest-weight module $W_{c,h}$ of weight $(c,h)$. The dual module, is the irreducible Verma module $V_{c,h}$. 

It has been shown~\cite{BB}, that
$$
\ker\hat{A}_{\infty}= W_{c,h}~.
$$

As next we would like to define an operator, acting on our probability space. Fix a domain $D_{A,B}$, and 
let $f:D'\rightarrow D$ be a  $({\cal F}_{D'_{A',B'}},{\cal F}_{D_{A,B}})$-measurable map. Define the action of the operator $R(f)$ by push-forward, i.e.
$$
R(f)\mu_{D_{A,B}}:=f_*\mu_{D'_{A',B'}}~.
$$

Now, let $D'\subset D $, such that $D\setminus D'$ is simply connected and $\overline{D\cap\complement D'}$ is compact. Then, any homeomorphism $f_{D'}:D'\rightarrow D$ can be considered as a partially defined function on $D$ itself. Nevertheless, we can consider 
$$
{f\Big|_{D'}}{}_*\,\mu_D
$$
Our definition of the sigma algebra is such that this partially defined function is still measurable.

If $\mu_D=:|h\rangle$, then the measure $\mu_D|\gamma_t$ is given by $R_{\gamma_t}|h\rangle$, where $R_{\gamma_t}:=R(f_{\gamma})_t$. 
The Radon-Nikodym derivative can be written as:
$$
\langle h|R_{\gamma_t} h\rangle:=\frac{\mu_D|\gamma_t}{\mu_D}=M_t({\gamma})
$$
Under a conformal automorphism $\rho\in\Aut_+({\cal O})$, which fixes two boundary points, and is without dilatation, we have for a conformally invariant probability measure $|h\rangle$:
$$
\langle h|R(\rho) h\rangle\equiv 1
$$

For a differentiable one-parameter family of biholomorphic maps $f_t$, $t\geq0$, such that 
$$
f_t:D\rightarrow f_t(D)=:D_t,
$$ 
and $f_0\equiv\id_D$, we have  
$$
\frac{d}{dt}\Big|_{t=0}\langle h|R(f^{-1}_t)h\rangle\equiv0~.
$$

Therefore the constant density $1$ can be seen as the ground state, which is annihilated by all $L_n$, $n\geq1$. This expresses the fact of conformal invariance of the measure. 

So we have
\begin{prop}[Virasoro action on measures]
The deformation space of conformally invariant measures on simple paths, arising from conditioning on a subarc, is given by a highest-weight representation of the Virasoro algebra in terms of differential operators, acting on polynomial densities.  
\end{prop}

Let us summarise what we have just gained. The SLE observable is the location of the tip on the boundary, after uniformisation. The physical observable is the partition function $Z_t$. The latter observable is defined as the value of the partition function of the slit domain, i.e.,  the boundary is enlarged by the segment $\gamma_t$. Both observables are random variables, as they depend on the particular realisation. 
SLE depends on $\kappa$, the physical observable $Z_t$ is characterised (in the simplest case) by two real variables $c$ and $h$. 
Therefore, the relation between conformally invariant measures on simple paths, derived from Statistical Mechanics, and SLE is given by the commutativity of the Diagram~\ref{SLEFactor}. 

Finally, the construction as a whole, can be seen, in principle, as a version of  Geometric Quantisation. On the base we have a transitive action of the Witt algebra, and on the total space, i.e., the twisted determinant line bundle, of the Virasoro algebra. In the representation as differential operators, factorisation requires the sections to be harmonic with respect to the lifted generator. This means that the partition function martingale is the Doob transform of the SLE process, or in the language of representation theory, is given by a degenerate highest-weight representation at level two. 
\section{Outlook}
We have seen the general framework which is underlying SLE. Further, we have given a rigourous link between conformally invariant measures, derived from CFT, with SLE. 

The canonical commutative diagram~\ref{SLEFactor}, which permits to couple the physical observable with the mathematical observable, is the key to generalisations, e.g. massive perturbations. 

Our demonstration for the special case, the disc, extends to the general situation, namely to arbitrary bordered Riemann surfaces. The necessary parts of that construction are gathered in the following diagram (for one marked point, and without denoting boundary components)~(cf.~\cite{F, FK, K, KS})
\noindent
\[
\begin{CD}
\label{trans_action}
\vir_{\C}@>\text{trans.}>>\pi^*\det^{\otimes c,h}@>>>\operatorname{Det}^{\otimes c,h}\\
@.@VVV@VVV\\
\witt@>\text{trans.}>>\hat{\mathfrak{M}}_{g,1}@>\text{Krichever}>>\Gr(H)\\ 
@.@VVV\\
@.\mathfrak{M}_{g,1}
\end{CD}
\]
where $\mathfrak{M}_{g,1}$ is the moduli stack of smooth pointed curves of genus $g>1$, and $\hat{\mathfrak{M}}_{g,1}$ is the moduli stack of triples $(X,p,z)$, where $(X,p)\in\mathfrak{M}_{g,1}$, and $z$ is a formal co-ordinate at $p$.
Then, $\hat{\mathfrak{M}}_{g,1}$
is an $\Aut({\cal O})$–bundle over $\mathfrak{M}_{g,1}$. The determinant line bundle $\pi^*\det^{\otimes c,h}$ is the pull-back of a tensor power of the corresponding determinant line bundle. Notably, the above can be applied, with some modifications, to Kac-Moody algebras~(cf.~\cite{FbZ}).

Also, the right object to obtain conformally invariant measures on loops on arbitrary Riemann surfaces, is again the Grassmannian of Sato-Segal-Wilson, combined with the line bundle on the loop space, coming from a sheaf of groupoids over the surface with connective structure.  

Finally, the connection with Integrable Systems should be now quite clear~(cf.~\cite{F09}).

\subsection*{Acknowledgements}
The author gratefully thanks the organisers of the MSJ-Seasonal Institute, held 2008 in Kyoto, for their hospitality and support.

\end{document}